\documentclass[10pt,journal,compsoc]{IEEEtran}
\ifCLASSOPTIONcompsoc
  \usepackage[nocompress]{cite}
\else
  \usepackage{cite}
\fi
\ifCLASSINFOpdf
\else
\fi
\IEEEoverridecommandlockouts
\usepackage{amsmath,amssymb,amsfonts}
\usepackage{algorithmic}
\usepackage{graphicx}
\usepackage{textcomp}
\usepackage[table]{xcolor} 
\usepackage{xcolor}
\usepackage{comment}
\usepackage{bm}
\usepackage[linesnumbered,algoruled,boxed,lined]{algorithm2e}
\usepackage{array}
\usepackage{diagbox}
\usepackage{makecell}
\usepackage{array}
\usepackage{booktabs} 
\usepackage{multirow}
\ifCLASSOPTIONcompsoc
\usepackage[caption=false, font=normalsize, labelfont=sf, textfont=sf]{subfig}
\else
\usepackage[caption=false, font=footnotesize]{subfig}
\fi

\def\BibTeX{{\rm B\kern-.05em{\sc i\kern-.025em b}\kern-.08em
    T\kern-.1667em\lower.7ex\hbox{E}\kern-.125emX}}
    
\begin{document}

\title{Optimizing Video Caching at the Edge:\\ A Hybrid Multi-Point Process Approach
\thanks{This work was supported by the National Natural Science Foundation of China under Grant U1911201, U2001209, ARC DE180100950, the project PCL Future Greater-Bay Area Network Facilities for Large-scale Experiments and Applications (LZC0019).}
}
\author{Xianzhi Zhang,
        Yipeng Zhou,~\IEEEmembership{Member,~IEEE,}
        Di Wu,~\IEEEmembership{Senior Member,~IEEE,}
        Miao Hu, ~\IEEEmembership{Member,~IEEE,}
        James Xi Zheng,
        Min Chen,~\IEEEmembership{Fellow,~IEEE,}
        and Song Guo,~\IEEEmembership{Fellow,~IEEE}
        
\IEEEcompsocitemizethanks{
\IEEEcompsocthanksitem Xianzhi Zhang, Di Wu and Miao Hu are with the Department of Computer Science, Sun Yat-sen University, Guangzhou, 510006, China, and Guangdong Key Laboratory of Big Data Analysis and Processing, Guangzhou, 510006, China. D. Wu is also with Peng Cheng Laboratory, Shenzhen 518000, China. (E-mail: zhangxzh9@mail2.sysu.edu.cn; \{wudi27, humiao5\}@mail.sysu.edu.cn.). 
\IEEEcompsocthanksitem Yipeng Zhou and James Xi Zheng are with the Department
of Computing, FSE, Macquarie University, Australia, 2122. Y. Zhou is also with Peng Cheng Laboratory, Shenzhen 518000, China. (E-mail: yipeng.zhou@mq.edu.au; james.zheng@mq.edu.au). 
\IEEEcompsocthanksitem Min Chen is with School of Computer Science and Technology, Huazhong University of Science and Technology, Wuhan 430074, China (E-mail: minchen@ieee.org).
\IEEEcompsocthanksitem Song Guo is with the Department of Computing, The Polytechnic University of Hong Kong, Hong Kong (E-mail: song.guo@polyu.edu.hk).

}
}

\markboth{Optimizing Video Caching at the Edge: A Hybrid Multi-Point Process Approach}%
{Zhang \MakeLowercase{\textit{et al.}}:Optimizing Video Caching at the Edge: A Hybrid Multi-Point Process Approach}

\IEEEtitleabstractindextext{
\begin{abstract}
It is always a challenging problem to deliver a huge volume of videos over the Internet. To meet the high bandwidth and stringent playback demand, one feasible solution is to cache video contents on edge servers based on predicted video popularity. Traditional caching algorithms (e.g., LRU, LFU) are too simple to capture the dynamics of video popularity, especially long-tailed videos. Recent learning-driven caching algorithms (e.g., DeepCache) show promising performance, however, such black-box approaches are lack of explainability and interpretability. Moreover, the parameter tuning requires a large number of historical  records, which are difficult to obtain for videos with low popularity. In this paper, we optimize video caching at the edge using a white-box approach, which is highly efficient and also completely explainable. To accurately capture the evolution of video popularity, we develop a mathematical model called \emph{HRS} model, which is the combination of multiple point processes, including Hawkes' self-exciting, reactive and self-correcting processes. The key advantage of the HRS model is its explainability, and much less number of model parameters. In addition, all its model parameters can be learned automatically through maximizing the Log-likelihood function constructed by past video request events. Next, we further design an online HRS-based video caching algorithm. To verify its effectiveness, we conduct a series of experiments using real video traces collected from Tencent Video, one of the largest online video providers in China. Experiment results demonstrate that our proposed algorithm outperforms the state-of-the-art algorithms, with  12.3\% improvement on average in terms of cache hit rate under realistic settings. 

\end{abstract}


}

\maketitle

\IEEEdisplaynontitleabstractindextext

\IEEEpeerreviewmaketitle

\IEEEraisesectionheading{\section{Introduction}\label{sec:introduction}}
\IEEEPARstart{D}{ue} to the fast growth of the online video market, the online video streaming service  has dominated the Internet traffic. It was forecasted by Cisco \cite{CiscoVNIMobile2019} that video streaming applications will take up the Internet traffic from 59\% in 2017  to 79\% in 2022. 
On one hand, online video providers need to stream HD (high definition) videos with  stringent playback requirements. On the other hand, both video population and  user population are growing  rapidly. Thereby, edge devices have been pervasively exploited by online video providers to cache videos so as to reduce the Internet traffic and improve the user Quality of Experience (QoE) \cite{Golrezaei2013,Gregori2016,Poularakis2014,Jiang2017,Yang2019,Muller2017}.  

We consider the video caching problem on  edge servers that  can provide video streaming services for users in a certain area  (\emph{e.g.}, a city).
If a video is cached, edge servers can stream  the video to users directly with a shorter response time. Nevertheless, requests for videos that are missed by edge servers have to be directed to remote servers, resulting in lower QoE. 
From video providers' perspective, the target is to maximize the cache hit rate of user video requests by properly caching videos on edge servers. 
Intuitively, videos to be requested most likely in the future should be cached on edge devices. It is common to leverage historical video requests to construct statistical models in order to predict videos' future request rates. 

 Briefly, there are two approaches to predict video popularity so as to make video caching decisions. The first approach is to predict video popularity based on empirical formulas. Classical LRU (Least Recently Used), LFU (Least Frequently Used) and their variants \cite{Jaleel2010,Famaey2013, Shafiq2014} are such video caching algorithms. 
 However, such approach is subject to the difficulty of choosing  parameter values. 
 For example, it is not easy to choose an appropriate time window size in LRU and LFU algorithms\cite{Jaleel2010,Ahmed2013}. 
 The second approach is learning-driven video caching algorithm. Typically, NN (neural network) models such as LSTM (Long Short-Term Memory)\cite{Feng2019,Narayanan2018} can be leveraged to predict video popularity so as to make right video caching decisions. 
 The strength of this approach is that model parameters can be automatically determined through learning historical request patterns, and thus such algorithms can achieve better video caching performance than classical algorithms \cite{Narayanan2018}. 
 Yet, such models  require a long training time, and are lack of explainability and interpretability.



In this paper, we aim at optimizing the performance of video caching on edge servers using a white-box approach. To this purpose, we develop a mathematical model called  \emph{HRS} model to capture the evolution process of video popularity. The HRS model is the combination of multiple point processes, including the Hawkes process\cite{Hawkes1971}, the reactive process\cite{Ertekin2015} and the self-correcting process\cite{Isham1979}. The point processes enable us to exploit timestamp information of historical video requests and time-varying kernels to predict future events. 

Specifically, the intuition behind the HRS model is as follows: with the Hawkes process, we can model the positive impact of the occurrence of a past event, and link future video request rates with the past video request events; with the reactive process, we can take the influence of negative events (e.g., the removal of a video from the recommendation list) into account;  with the self-correcting process, we can restrict the growth of video popularity. 
Compared to NN(neural network)-based models, our proposed HRS model is completely explainable and interpretable. Moreover, the number of parameters of HRS model is much less than that of NN-based models. 

 

In summary, our main contributions in this paper can be summarized as below:
\begin{itemize}
\item We develop a hybrid multi-point process model called \emph{HRS} to accurately predict the evolution of video popularity (i.e., video request rate). Different from NN-based model, our HRS model is completely explainable and interpretable.  The HRS model can link future video request rates with both past video request events and negative events, and take the characteristics of edge servers into account. 

\item We propose an online video caching algorithm for edge servers based on the HRS model. Due to much less model parameters, the algorithm has very low computation complexity. All parameters of the HRS model can be determined automatically by maximizing the Log-likelihood function. Thus, the algorithm can be executed frequently to update cached videos timely according to the dynamics of video popularity.  
\item We conduct extensive real trace-driven experiments to validate the effectiveness of our proposed algorithm. The video request traces are collected from Tencent Video, one of the largest online video providers in China. The experimental results show that the HRS-based online caching algorithm can achieve the highest cache hit rate with 12.3\% improvement than the best baseline algorithm in terms of cache hit rate. In particular, the improvement is over 24\% when the caching capacity is very limited.  In addition, the execution time of our algorithm is much lower than that of NN-based caching algorithms. 

\end{itemize}

The rest of the paper is organized as follows. We first provide an introduction of preliminary knowledge in the next section which is followed by the description of the HRS model in Sec \ref{HRS model}. Notably, the new online caching algorithm is proposed in Sec. \ref{Online HRS Algorithm}. The experimental results are presented in Sec \ref{Evaluation}; while the related works in this area are discussed in Sec \ref{Related Work} before we finally conclude our paper.

\section{Preliminary}\label{sec:Preliminary}
\subsection{Video Caching On Edge}
\begin{figure}[!t]
    \centering
    \includegraphics[width=3.5 in]{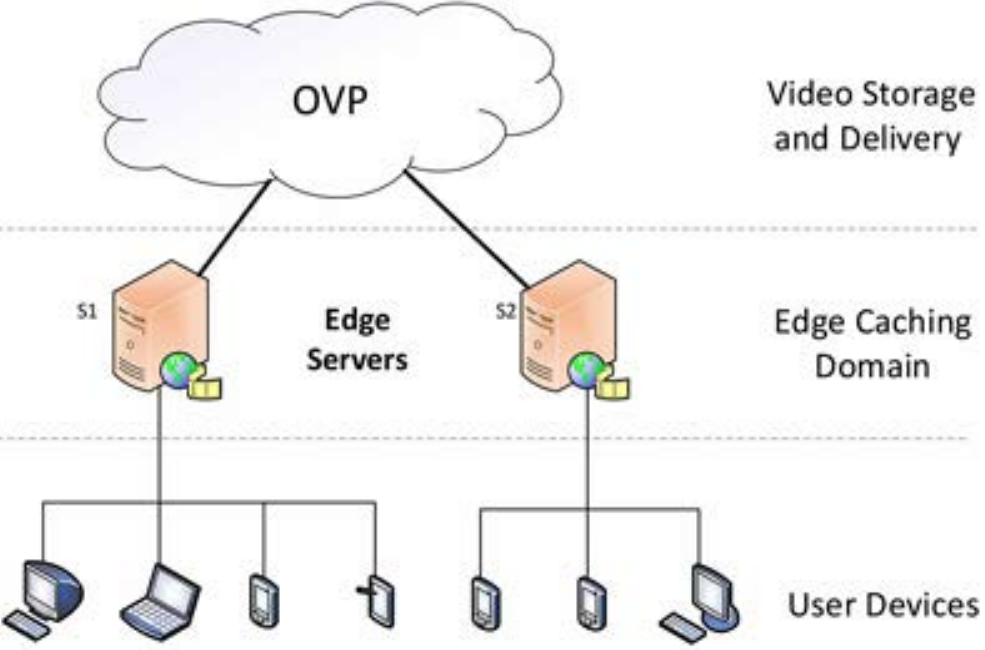}
    \caption{The system architecture of the video caching system with multiple edge servers.}
    \label{EDGECACHING}
    \end{figure}

We consider a video caching system with multiple edge servers. The system architecture is illustrated in Fig.~\ref{EDGECACHING}. Online video providers (OVP)  store a complete set of videos with population $C$. A number of edge servers are deployed in the proximity of end users. Each edge server can exclusively cover users in a certain area.
Since edge servers are closer to end users,  serving users with edge servers can not only reduce the Internet traffic but also improve the user QoE \cite{Yang2019}. 
The problem is to predict the video request rates on each edge server in the future so that the most popular videos can be cached in time by the edge server. 

Our objective is to maximize the cache hit rate on each edge server, which is defined as the number of requests served by the edge server divided by the total number of requests. 
This objective can be transformed to maximize the  prediction accuracy of future video request rates. 

Without loss of generality, we consider the video caching problem for a particular edge server, which can store $S$ videos with $S<C$. 
To simplify our analysis, we assume that all videos are of the same size. In real systems, videos with different sizes can be split into blocks of the same size. This simplification can significantly reduce the implementation cost in real video caching systems  \cite{Poularakis2014,Shanmugam2013}.
With this simplification, it is  apparent that the top $S$ most frequently requested videos should be cached on each edge server, and our main task is to predict which $S$ videos will be most popular in the future.  For convenience, major notations used in this paper are summarized in Table \ref{Major Notations}.


\begin{table*}[!htbp]
\renewcommand\arraystretch{1.2}
\caption{Notations used in the paper}
\begin{center}
\rowcolors{2}{white}{gray!25} 
\begin{tabular}{p{2.7cm}<{\centering} m{14.5cm}}
\toprule
Notation   & \multicolumn{1}{c}{Description}\\ 
\midrule
 $C$ &  The total number     of videos stored in OVP.   \\
 $S$ &  The total number of videos cached on an edge server.  \\
 $K$ & The total  number of  request records.\\
 $\mathcal{E}$ & The event set of all requested records.\\
 $\varepsilon^t$ / $\zeta^t$   &  The timestamp set of all requested records / negative events before time $t$. \\
 $\varepsilon^{t}_i$  / $\zeta^t_i$  & The timestamp set of requested records / negative events of video $i$  before time $t$. \\
 $\tau$ / $\tau'$ & The timestamp of any request record / negative event.\\
 $\lambda_i\left( t \right)$ & The conditional intensity function of video $i$ at time $t$.\\
 $\tilde{\lambda}_{i}(t)$& The estimation of $\lambda_{i}(t)$, which is defined by Eq.~\eqref{HRS intensity function} for HRS.\\ 
  $\hat{\lambda}_{i}(t)$& The positive form of $\tilde{\lambda}_{i}(t)$ adjusted by $g(x) = s \log(1 + \exp(x/s))$.\\ 
 $\beta_i$ & The bias of intensity function for HRS associated with video $i$.\\
 $\omega_i$ / $\alpha_i$ / $\gamma_i$& The parameter of SE / SC / SR term in HRS associated with video $i$ respectively.\\
 $k_0(t-\tau)$ / $k_1(t-\tau')$ & The exponential kernel function of SE / SR term reflecting the influence of past event defined by Eq.~\eqref{EQ:KernelFun}.\\
 $\delta_0$ / $\delta_1$ & The decay parameter of $k_0(t-\tau)$ / $k_1(t-\tau')$, which can be determined through cross validation in experiments.\\
 $T$ &  The entire observation period time for the computation of likelihood function.   \\
 $\bm{\theta}$& The parameters matrix of a point process and $\bm{\theta}=\left[\ \bm{\beta}^{\intercal},\bm{\omega}^{\intercal},\bm{\alpha}^{\intercal},\bm{\gamma}^{\intercal}\ \right]$ in HRS model. \\
  $\theta_i^{(j)}$& The substitute of any parameter associated with video $i$ after $j$ iterations in gradient descent algorithm, such as $\beta_i^{(j)}$.\\ 
 $ll(\bm{\theta})$ & The Log-likelihood function of point process, which is defined by Eq.\eqref{system likelihood function} and Eq.\eqref{system likelihood function rewrite} for HRS.\\
 $\bar{ll}(\bm{\theta})$ & The evaluated Log-likelihood function of HRS, which is estimated by Monte Carlo method and defined by Eq.~\eqref{system likelihood function evaluated}.\\ 
  $\rho_{\beta}$ / $\rho_{\omega}$ / $\rho_{\alpha}$ / $\rho_{\gamma}$ & The regularization parameter of $\beta_i$ / $\omega_i$ / $\alpha_i$ / $\gamma_i$, which can be determined through cross validation in experiments. \\
 $M$ & The number of sample times in Monte Carlo method. \\
 $\mathbf{t}^{(m)}$ & The timestamp of the $m$-th sample in Monte Carlo method.\\
 $\Phi_i(t)$ / $\Psi_i(t)$ / $\Gamma_i(t)$ & The kernel function of HRS defined by Eq.~\eqref{EQ:CoreTermComp}.\\
 $\Delta t$ & The time interval of online kernel functions update. \\ 
 $\Delta T$ & The time interval of online parameter update. \\ 
 $\Delta M$ & The number of sample times of online parameter update.\\
 $k_{th}$ & The threshold to truncate the sum of $k_0$ of online parameter update.\\
\bottomrule
\end{tabular}
\end{center}
\label{Major Notations}
\end{table*}   

We first define the event sets as follows. 
All past video requests are denoted by an event set $\mathcal{E} = \{e_1, e_2, ...,  e_K\}$. Let $\varepsilon = \{\tau_1,\tau_2,...,\tau_K\}$ denote the occurrence time of all past events, $\tau_1<\tau_2<,\dots, <\tau_K$. In other words, events are recorded according to their  occurrence time points. Each event in the set $\mathcal{E}$ is a tuple, \emph{i.e.}, $e =\left \langle\, \tau, i \,\right \rangle$, where $\tau $ is the request time and  $i$ is the video index. Besides, we define the set $\varepsilon^{t}_i$ as the timestamp set for historical events of video $i$  before time $t$ and $\varepsilon^{t}= \cup_{\forall i} \varepsilon^{t}_{i}$.

\subsection{{Point Process}}

Point process is  a family of models which are generated from on individual events to capture the temporal dynamics of event sequences via the conditional intensity function\cite{Daley2008}. 
We employ the point process models to predict the future request rate given the historical request events of a particular video. 
In general, the predicted request rate of video $i$ can be  defined as $\lambda_i\left( t | \varepsilon^t_i \right)$, where $\varepsilon^t_i$ is the  timestamp set of historical requested records of video $i$ before time $t$.  
The conditional intensity function represents the expected instantaneous request rate of video $i$ at time $t$. 
In the rest of the paper, we use $\lambda_i(t)$ to represent $ \lambda_i\left( t | \varepsilon^t_i \right)$.

We first introduce three typical point process models before we specify the expression of $\lambda_i(t)$, 

\subsubsection{Hawkes Process (HP)} HP \cite{Hawkes1971} is  also called self-exciting process. 
For HP, the occurrence of a past event will positively affect the arrival rate of the same type of events in the future. 
Given the timestamp set of historical events $\varepsilon_i^t$ , the arrival rate of such events at time $t$ can be predicted according to the following formula:  
\begin{equation}
        \lambda_i(t) = \beta_i + \sum_{\forall \tau\in\varepsilon_i^t}\phi_i(t-\tau).
        \label{general intensity}
\end{equation}

Here,  $\beta_i$ is the deterministic base rate and $\phi_i(t-\tau)$ is the kernel function to reflect the influence of the past event at time $\tau$ on the arrival rate of the same type at time $t$. Moreover, $\phi_i(t-\tau)$ should be a monotonically decreasing function with $t-\tau$ in that more recent events have higher positive influence on the future request rate. 
    

\subsubsection{Reactive Process}

The reactive process\cite{Ertekin2015} is an extension of HP by linking the future event with more than one types of events. HP only considers the influence of positive events, while the reactive process models both exciting and restraining effects of historical events on instantaneous intensity. The future rate can be given by:
    \begin{equation}
        {\lambda}_i(t) =\beta_i + \sum_{\forall \tau \in \varepsilon_i^t}\phi_i^{exc}(t-\tau) - \sum_{\forall \tau' \in \zeta^t_i}\phi_i^{res}(t-\tau').
    \end{equation}
    
Here $\varepsilon_i^t$ denotes the same timestamp set of positive events as that in HP while $\zeta^t_i$ represents the timestamp set of negative events restraining the intensity function of video $i$ before time $t$. $\phi_i^{exc}$ and $\phi_i^{res}$ are kernel functions to reflect the influence of positive and negative events respectively. 



\subsubsection{Self-correcting Process}    
Compared to the Hawkes process and reactive process, the intensity function of the self-correcting process is more stable over time \cite{Isham1979}. Once an event occurs, the intensity function will be reduced by a factor $e^{-\alpha_i}$. Here $\alpha_i$ is a parameter representing the ratio of correcting. Mathematically, the intensity function can be given by
    \begin{equation}
    \begin{split}
        \lambda_i(t) = \exp(\mu_i t - \alpha_i{N}_{i}(t)),
       \end{split} 
    \end{equation}
where $\mu_i$ is the rate of the steadily increase of intensity function and ${N}_{i}(t)$  is the number of historical requests of video $i$ until time $t$. Generally, the time series of events based on the self-correcting process are more uniform than those based on other processes such as HP. 
    

\subsubsection{Log-likelihood Function of Point Process} 

To apply the point process models to predict video request rates, we need to determine the parameters   (denoted by $\bm{\theta}$) defined in each  process, such as $\mu$ and  $\alpha$ in the self-correcting process.
An effective approach to determine these parameters is to maximize the likelihood of the occurrence of a target event set in the entire  observation time $(\,0,T\,]$, \emph{i.e.}, $\mathcal{E}^T$.

Given the overall intensity function $\lambda(t) = \sum_{\forall i}\lambda_i(t)$ and the  occurrence time  $\tau_{l}$ of the last event  in  the historical event time set $\varepsilon^t$, the probability that no event $i$ occurs in the period $[\,\tau_{l}, t\,)$ is  $P\left(\textit{no event in $[\,\tau_{l}, t\,)$} \ \big| \ \varepsilon^t\right) = \exp \left(-\int_{\tau_{l}}^{t} \lambda(x) \mathrm{d} x\right)$. Thus, the probability density that an event $i$ occurs at time $t$ is given by:
	\begin{align}
	\begin{split}
        P\left(t,i \ \big| \ \varepsilon^t\right)&=\lambda_i(t) \exp \left(-\int_{\tau_{l}}^{t} \lambda(x) \mathrm{d} x\right).
    \end{split}
    \end{align}
    
The detailed derivation can be found in \cite{Rizoiu2017}. Given the event set $\mathcal{E}^T = \{e_1, \dots, e_{K}\}$, where $e_k = \left \langle \tau_k, i_k \right \rangle $ and $K$ is the total number of events during the time interval $(\,0,T\,]$, it is easy to derive the likelihood function for a given event set using Eq.~(\ref{L}). 
    \begin{equation}
    \begin{split}
        L \left( \, {\bm{\theta}} \, \big| \, \mathcal{E}^T \, \right)
        &= \prod_{ k=1}^{K}\lambda_{i_k}(\tau_k)
          \exp \left(-\int_{\tau_{k-1}}^{\tau_k} \lambda(t) \mathrm{d} t\right)\\ 
          &\times \exp \left( -\int_{\tau_k}^{T} \lambda(t) \mathrm{d} t\right) ,\\
        &= \prod_{\forall i}\prod_{\tau \in \varepsilon^T_i}  \lambda_i(\tau) \times \exp \left( -\int_{0}^{T} \lambda(t) \mathrm{d} t\right) .
    \label{L}
	\end{split}
    \end{equation}

For convenience, we let $\tau_0 = 0$ and align all time series for different videos $i$ to the same initial point $\tau_0 = 0$. In practice, it is difficult to manipulate the likelihood function. Equivalently, we can optimize the Log-likelihood function, which is defined as:
    \begin{equation}
    \begin{split}
        ll \left( \, \bm{\theta} \, \big| \, \varepsilon^T \, \right) =   \sum_{\forall i} \sum_{\forall \tau \in \varepsilon^T_i} \log\left( \lambda_i(\tau)\right) - \int_{0}^{T} \lambda(t) \mathrm{d} t, 
    \label{EQ:logL}
	\end{split}
    \end{equation}
    where $\varepsilon^{T}$ is the timestamp set of target events in the entire observation time interval $(\,0,T\,]$.

\section{HRS Model \label{HRS model}}

In this section, we describe the HRS model which is the combination of three point process models introduced in the last section. 

\subsection{{Intensity Function Design}}

First of all, we explain the intuition behind the HRS model, and the reasons why we take three types of point processes into account.
\begin{itemize}
    \item \emph{Self-exciting}: If a video has attracted a number of user requests recently, each request can impose some positive influence on the future request rate of the same video. This is the self-exciting process depicted by the Hawkes process. It can well capture videos that are becoming more and more popular. 
    \item \emph{Reactive process}: Different from the positive influence of historical request events, there also exist events that impose negative influence on the future request rates. For example, the popularity of a video may sharply drop down if the video is removed from the recommendation list.  Such negative events can be modeled by the reactive process.
    \item \emph{Self-correcting}: 
    The user population covered by an edge server is limited. Thus, if users do not repeatedly request videos, a video can not stay popular forever. In fact, users seldom replay videos they have watched \cite{Wu2016}. It is expected that the popularity of a video will diminish with time after a majority of users have requested it. The restriction of the limited user population can be captured by the self-correcting process. 
\end{itemize}

Based on the above discussion, we can see that the evolution of video popularity is a very complicated process. 
It is difficult to precisely model the video request rate by merely utilizing only one particular kind of point process.
In view of that, we propose to construct the video request intensity function by combing three types of point processes together. The proposed request rate intensity function is presented by:
	\begin{equation}
	\begin{split}
		\tilde{\lambda}_{i}(t)=&
			\underbrace{\beta_{i}}_{\text{bias}}+ \overbrace{\omega_{i}  \sum_{\tau \in \varepsilon_{i}^{t}}  k_0(t-\tau) }^{\text{self-exciting}} \underbrace{\exp[-\alpha_{i} {N}_{i}(\tau)] }_{\text{self-correcting}}
			\\
	        & - \underbrace{\gamma_{i}\sum_{\tau' \in \zeta_i^{t}}  k_{1}(t-\tau')}_{\text{self-restraining}},
	 \label{HRS intensity function}
	\end{split}
	\end{equation}
where $\lambda_{i}(t)$ is the true request rate of video $i$ at time $t$, and $\tilde{\lambda}_{i}(t)$ is the estimation of $\lambda_{i}(t)$. Each term in Eq.~(\ref{HRS intensity function}) is explained as follows.


\begin{itemize}
    \item The first term $\beta_{i}$ is the bias of the intensity function for video content $i$ with positive value ($\beta_i>0$).
    \item As we have marked in Eq.~\eqref{HRS intensity function}, $\omega_{i}  \sum_{\tau \in \varepsilon_{i}^{t}}  k_0(t-\tau)$ is the SE (self-exciting) term. It is designed based on the Hawkes process and $\omega_{i}   k_0(t-\tau)$ is the positive influence imposed by the video request event at time $\tau$. $ k_0(t-\tau)$ is a kernel function to be specified later and $\omega_i$ is a parameter to be learned.
    \item The second term $\gamma_{i}\sum_{\tau' \in \zeta_i^{t}}  k_{1}(t-\tau')$, which is called SR (self-restraining) term, captures the influence of negative events such as the event that a video is removed from the recommendation list. This SR term is designed based on the reactive process.  $\zeta_i^{t}$ is the set including the timestamp of all negative events in the period $[0,t)$. Similar to modeling the influence of positive events, $k_1(t-\tau') $ is the kernel function to account for the influence of a negative event. $\gamma_i$ is a parameter to be learned.
    \item The last term $\exp[-\alpha_{i} {N}_{i}(\tau)]$ is the SC (self-correcting) term and  ${N}_{i}(\tau)$ is the number of historical requests of video $i$ until time $\tau$. The implication is that the influence of a positive event will be smaller if more users have watched the video $i$. For example, if there are two movies: A and B.  Movie A  has been watched by 99\% users, but Movie B is a new one watched by only 1\% users. Then, the influence of a request event for Movie A or B should be very different. 

\end{itemize}
    
For point process models, it is common to adopt exponential kernel functions to quantify the influence of historical events\cite{Rizoiu2017,Hawkes1971,Daley2008}. Thus, in the HRS model, the kernel functions  $k_0(t-\tau)$ and  $k_1(t-\tau')$ are set as:
\begin{equation}
\begin{subequations}
    \begin{aligned}
        &k_0(t-\tau) = exp[-\delta_0 (t-\tau)],\\
        &k_1(t-\tau) = exp[-\delta_1 (t-\tau')].
    \end{aligned}
    \label{EQ:KernelFun}
\end{subequations}
\end{equation}
Here $\delta_0> 0$ and $\delta_1> 0$ are two hyper-parameters, which can be determined empirically through cross validation.
From Eq.~\eqref{EQ:KernelFun}, we can observe that kernel functions decay with  $t-\tau$, implying that the influence gradually diminishes with time. 

 
By considering the reality of video request rates, we need to impose restrictions on the intensity function defined in Eq.~\eqref{HRS intensity function}. 
\begin{itemize}
    \item The video request rate is non-negative. However, due to the SR term, it is not guaranteed that Eq.~\eqref{HRS intensity function} always yields a non-negative request rate. Besides, the  Log-likelihood function requires that the video request rate must be positive. Thus, we define 
    \begin{equation}
		\hat{\lambda}_{i}(t)=s \log\left(1 + \exp(\tilde{\lambda}_{i}(t)/s)\right).
		\label{HRS intensity function positive}
	\end{equation}
	Here $s$ is a small positive constant number. We utilize the property that the function  $g(x) = s \log(1 + \exp(x/s)) \approx \max\{0,x\}$, \emph{i.e.}, the ReLU function, as $s\to 0$ \cite{Mei2017a}. 
	\item All parameters $\beta_i$, $\omega_i$, $\alpha_i$ and $\gamma_i$ should be positive numbers to correctly quantify the influence of each term. 
\end{itemize}

The intensity function $\hat{\lambda}_{i}(t)$ is the final estimation form of the  request intensity $\lambda_{i}^t$ for video $i$ at time $t$.

\subsection{{ Maximizing Log-likelihood Function\label{offline training method}}}

Given the event sets $ \varepsilon^{T}$ and $\zeta^T$, and the parameters  $\bm{\theta}=\left[\ \bm{\beta}^{\intercal},\bm{\omega}^{\intercal},\bm{\alpha}^{\intercal},\bm{\gamma}^{\intercal}\ \right]$,  the Log-likelihood function  is defined according to Eq.~\eqref{EQ:logL} as:
	\begin{equation}
	ll\left( \, \bm{\theta} \, \big| \, \varepsilon^{T},\zeta^{T} \, \right)
	=\sum_{\forall i}\sum_{\tau \in \varepsilon_{i}^{T}} \log \hat{\lambda}_{i}(\tau)  - \sum_{\forall i}\int_{0}^{T} \hat{\lambda}_{i}(t) \, \mathrm{d} t. \label{system likelihood function}
	\end{equation}

In the rest of this work, we use the shorthand notation $ll\left( \, \bm{\theta} \,\right) =: ll\left( \, \bm{\theta} \, \big| \, \varepsilon^{T},\zeta^{T} \, \right)$ if event sets are clear in the context. In addition, to simplify our notations, let  $\mathcal{I}_i = \int_0^T\hat{\lambda}_{i}(t) \, \mathrm{d} t $ represent the integral term in the Log-likelihood function of video $i$. Then,  the Log-likelihood function can be rewritten as:
	\begin{equation}
	\begin{split}
	ll\left( \, \bm{\theta} \,\right)  = & \sum_{\forall i} \sum_{\forall\tau\in\varepsilon_{i}^{T}} \log \hat{\lambda}_{i}(\tau)- \sum_{\forall i}\mathcal{I}_i.
	\label{system likelihood function rewrite}
	\end{split}
	\end{equation}	

The challenge to maximize the Log-likelihood function lies in the difficulty to derive $\mathcal{I}_i$ due to the complication of Eq.~\eqref{HRS intensity function positive}. Thus, we resort to the  Monte Carlo estimator to derive  $\mathcal{I}_i$ approximately. 

We briefly introduce the  Monte Carlo estimator as follows. Given a function $f(x)$, $M$ samples of $x$ can be uniformly and randomly selected from the domain of $x$, say $(a, b)$. Then, the integral of $f(x)$ can be calculated as:
	\begin{equation}
		\begin{aligned}
		\mathcal{I}&  = \int_{a}^{b}f(x) \, \mathrm{d} t =\mathbb{E}\left[\bar{\mathcal{I}}^{M}\right],\\
		\text{ s.t. } & \bar{\mathcal{I}}^{M} = \frac{1}{M} \sum_{m=1}^{M} \frac{f(\mathbf{x}^{(m)})}{p(\mathbf{x}^{(m)})},\\\
		& \mathbf{x}^{(m)} \sim \text{U}(a,b),\\
		& p(\mathbf{x}^{(m)}) = \frac{1}{b-a},
		\label{Monte Carlo trick}
		\end{aligned}
	\end{equation}
where $a$ and $b$ are lower and upper limit points of the integral function respectively. The expected value of the integral term, \emph{i.e.}, $\mathcal{I} = \mathbb{E}\left[\bar{\mathcal{I}}^{M}\right]$, can be approximately performed by the average of $f(\mathbf{x}^{(m)})/p(\mathbf{x}^{(m)})$. 

Suppose the integral range of  $\mathcal{I}_i$ is from $0$ to $T$, we can apply the Monte Carlo estimator to evaluate $\mathcal{I}_i$  as follows:
	\begin{equation}
	\begin{aligned}
		\mathcal{I}_i &\approx\frac{T}{M} \sum_{n=1}^{M} \hat{\lambda}_i\left(\mathbf{t}^{(m)}\right),\\
		\text { s.t. }&  \mathbf{t}^{(m)} \sim \text{U}(0,T).
		\label{estimate integral}
	\end{aligned}
	\end{equation}	
	
The Log-likelihood function can be approximately evaluated by:
	\begin{equation}
	\begin{aligned}
	\bar{ll} \left( \, \bm{\theta} \,  \right)
	&= \sum_{\forall i} \sum_{\forall\tau\in\varepsilon_{i}^{T}} \log \hat{\lambda}_{i}(\tau)-\sum_{\forall i}   \bar{\mathcal{I}}_i^{M},
	\\
	&= \sum_{\forall i} \sum_{\forall\tau\in\varepsilon_{i}^{T}} \log \hat{\lambda}_{i}(\tau)- \frac{T}{M}\sum_{\forall i}  \sum_{m=1}^{M}   \hat{\lambda}_i\left(\mathbf{t}^{(m)}\right),
	\\
     & \qquad \qquad \text { s.t. } \quad  \mathbf{t}^{(m)} \sim \text{U}(0,T).
    \label{system likelihood function evaluated}
	\end{aligned}
	\end{equation}

Equivalently, we can minimize the negated Log-likelihood function. By involving the regularization terms, our problem can be formally defined as:
	\begin{equation}
	\begin{aligned}
	\min\limits_{\bm{\theta}} \ \mathcal{L} = - \, \bar{ll}\left( \, \bm{\theta} \,  \right) &+\rho_{\beta}\|\bm{\beta}\|_{2}+\rho_{\omega}\|\bm{\omega}\|_{2} \\
	&+ \rho_{\alpha}\|\bm{\alpha}\|_{2}+\rho_{\gamma}\|\bm{\gamma}\|_{2},
	\\
	\text { s.t. } \quad {\beta_i}, {\omega_i}, {\alpha_i},& {\gamma_i}>0, \text { for } \forall i,
	\end{aligned}
	\label{system LOSS function}
	\end{equation}
where $\rho_{\beta},\rho_{\omega},\rho_{\alpha}$ and $\rho_{\gamma}$ are regularization parameters. 

Since the Log-likelihood in Eq.~(\ref{system likelihood function evaluated}) is a convex function, we can solve  Eq.~(\ref{system LOSS function}) by using the Gradient Descent (GD) algorithm \cite{Daley2008,Rizoiu2017}. 
By differentiating $\bar{ll}\left( \, \bm{\theta} \right)$ with respect to each parameter $\theta_i  \in \bm{\theta}$,\footnote{$\theta_i$ is a parameter associated with video $i$, such as $\beta_i$.} one can derive the following results:
	
	\begin{align}
	\begin{split}
	\frac{\partial \, \bar{ll} \left( \, \bm{\theta} \,  \right)}{\partial \theta_i}= &  \sum_{\tau \in \varepsilon_{i}^{T}} \frac{1}{\hat{\lambda}_{i}(\tau)}  \frac{\partial \hat{\lambda}_i(\tau)}{\partial \theta_i}-\frac{T}{M}\sum_{m=1}^M \frac{\partial \hat{\lambda}_i(\mathbf{t}^{(m)})}{\partial \theta_i}. 
	\label{Eq:l dev theta_i}
	\end{split}
	\end{align}
	
According to Eq.~(\ref{Eq:l dev theta_i}),  we need to derive  $\partial \hat{\lambda}_i(t)\,/\,\partial \theta_i$ in order to obtain the gradient of $\bar{ll}\left( \, \bm{\theta} \right)$. Thus, by differentiating $\hat{\lambda}_i(t)$ shown in Eq.~\eqref{HRS intensity function positive} with respect to each parameter, we can obtain  Eqs.~(\ref{lambda dev betai})-(\ref{lambda dev gammai}) as follows:
	\begin{align}
		\begin{split}
			\frac{\partial \hat{\lambda}_{i}(t)}{\partial \beta_{i}}= & \frac{\partial \hat{\lambda}_{i}(t)}{\partial \tilde{\lambda}_{i}(t)}, \label{lambda dev betai}
		\end{split}
		\\
		\begin{split}
			\frac{\partial \hat{\lambda}_{i}(t)}{\partial \alpha_{i}} = 
			& - \frac{\partial \hat{\lambda}_{i}(t)}{\partial \tilde{\lambda}_{i}(t)} \omega_{i}\sum_{\tau \in \varepsilon_{i}^{t}} k_{0}\left(t-\tau\right) {N}_{i}(\tau)\exp[- \alpha_{i} {N}_{i}(\tau)] , \label{lambda dev alphai}
		\end{split}
		\\
		\begin{split}
			\frac{\partial \hat{\lambda}_{i}(t)}{\partial \omega_{i}}=
			&  \frac{\partial \hat{\lambda}_{i}(t)}{\partial \tilde{\lambda}_{i}(t)}\sum_{\tau \in \varepsilon_{i}^{t}}  k_0(t-\tau)  \exp[-\alpha_{i} N_{i}(\tau)]\label{lambda dev omegai},
		\end{split}
		\\
		\begin{split}
			\frac{\partial \hat{\lambda}_{i}(t)}{\partial \gamma_{i}}=
			&-\frac{\partial \hat{\lambda}_{i}(t)}{\partial \tilde{\lambda}_{i}(t)}\sum_{\tau' \in \zeta_i^{t}}  k_{1}(t-\tau')
		\end{split}. \label{lambda dev gammai}
	\end{align}
	
	Here $\partial \hat{\lambda}_{i}(t)/\partial \tilde{\lambda}_{i}(t)$ can be calculated by Eq.~(\ref{lambda dev lambda}), which gives
	\begin{equation}
        \frac{\partial \hat{\lambda}_{i}(t)}{\partial \tilde{\lambda}_{i}(t)}= \frac{\exp(\tilde{\lambda}_{i}(t)/s)}{ 1 + \exp(\tilde{\lambda}_{i}(t)/s)}.
        \label{lambda dev lambda}
    \end{equation}



	 By integrating Eqs.~(\ref{lambda dev betai})-(\ref{lambda dev lambda}) with Eq.~(\ref{Eq:l dev theta_i}), we can obtain the gradient expression of $\bar{ll}\left( \, \bm{\theta} \right)$. Then, let $\theta^{(j)}_i$ represent any parameter to be determined after $j$ iterations. We update $\theta_i$ according to:
	\begin{equation}\theta^{(j+1)}_i \leftarrow \theta^{(j)}_i+\eta \left(- \frac{\partial \, \bar{ll} \left( \, \bm{\theta} \, \right)}{\partial \, \theta^{(j)}_i} + \rho_{\theta_i} \theta^{(j)}_i\right),
	\label{update rule}
	\end{equation}
where $\rho_{\theta_i}$ is the regularization parameter associated with the parameter $\theta_i$ and $\eta$ is the  learning rate depending on the GD algorithm. In our work, to adhere to Eq.~(\ref{update rule}) and ensure all parameters are within the boundaries specified in Eq.~(\ref{system LOSS function}), we adopt the L-BFGS \cite{Byrd1995} algorithm to minimize the objective function. 

\subsection{{Analysis of Computation Complexity}}

The computation complexity for a round of iteration to determine the parameters in the HRS model is $O(|\varepsilon^T|+|\zeta^T|+CM)$, where $|\varepsilon^T|$ and $|\zeta^T|$ are the total numbers of positive and negative events respectively, $C$ is the total number of videos and $M$ is the  Monte Carlo sampling number of each video. 

We analyze the detailed computation complexity as follows. 
To minimize the objective function in Eq.~\eqref{system LOSS function}, it is necessary to  compute  gradients according to Eq.~\eqref{lambda dev betai}-\eqref{lambda dev gammai}. To ease our discussion, we define three functions as below:
    \begin{subequations}
	\begin{align}
            &\Phi_i(t) = \sum_{\tau \in \varepsilon_{i}^{t}}  k_0(t-\tau)  \exp[-\alpha_{i} N_{i}(\tau)] \label{EQ:PhiCoreTermComp}\\
            &\Psi_i(t) = \sum_{\tau' \in \zeta_i^{t}}  k_{1}(t-\tau')\label{EQ:PsiCoreTermComp}\\
            &\Gamma_i(t)  =  \sum_{\tau \in \varepsilon_{i}^{t}} k_{0}\left(t-\tau\right) N_{i}(\tau)\exp[- \alpha_{i} N_{i}(\tau)]\label{EQ:GammaCoreTermComp}
    \end{align}
    \label{EQ:CoreTermComp}
    \end{subequations}

We need to compute $\varepsilon^T_i$ for different $\Phi_i(t)$'s and $\Gamma_i(t)$'s, and $\zeta^T_i$ for $\Psi_i(t)$'s. Specifically, for a particular event with occurrence time $\tau$ and its previous event with occurrence time $\tau_l$, we can compute $\Phi_i(\tau) = \Phi_i(\tau_l)exp[-\delta_0(\tau-\tau_l)]+ \exp[-\alpha_{i} N_{i}(\tau)]$.\footnote{Here, we utilize the property that $k_0(0) = 1$}  Thus, the computation complexity is $O(1)$   to compute $\Phi_i(t)$ for each event and the complexity is $O(|\varepsilon_i^T|)$ to complete the computation for all video $i$ events. Recalling that $\varepsilon^T= \cup_{\forall i} \varepsilon^T_{i}$, the whole computation complexity for all $\Phi_i(t)$'s is $O(|\varepsilon^T|)$. Similarly, the computation complexity is $O(|\varepsilon^T|)$/$O(|\zeta^T|)$ to compute all $\Psi_i(t)$'s/$\Gamma_i(t)$'s. 

With the above computations, the complexity to compute the first term of Eq.~(\ref{Eq:l dev theta_i}) is $O(|\varepsilon^T|)$.
 For the Monte Carlo estimator, it needs to sample $M$ times for each video.  
 Thus, there are totally $CM$ samples in a round of iteration for $C$ videos.  Besides, given the sampled time point $\mathbf{t}^{(m)}$, the complexity is $O(1)$ to compute all gradients. By wrapping up our analysis, the overall computation complexity for each iteration is $O(|\varepsilon^T|+|\zeta^T|+CM)$.\footnote{In fact, this complexity is an upper bound since we can complete the computation of $\Psi_i(t)$ in the first iteration without the necessity to update it in  subsequent iterations. } 


\section{Online HRS Based Video Caching Algorithm\label{Online HRS Algorithm}}
The complexity $O(|\varepsilon^T|+|\zeta^T|+CM)$  is not high if the training algorithm  is only executed once. 
Yet, the online video system is dynamic because  user interests can change over time rapidly, and fresh videos (or users) enter the online video system continuously.
The computation load will be too heavy if we need to update the predicted video request rates very frequently. 
Thus, in this section, we propose an online video caching algorithm based on the HRS model. 
The online algorithm can update the predicted video request rates by minor computation with incremental events since the last update.  


The online video caching algorithm needs to cope with two kinds of changes. The first one is the kernel function update. Given the HRS model parameters, the video request rates predicted according to Eq.~\eqref{HRS intensity function} should be updated according to the latest events. The user interest can be very dynamic. For example, users may prefer News videos in the morning, but Movie and TV videos in the evening. Thus, the predicted request rates should be updated instantly and frequently in accordance with the latest events. 
The second one is the parameter update. The model parameters such as $\omega_i$ and $\gamma_i$ capture the influence weight of each term in the HRS model. In a long term, due to the change of users and videos, model parameters should be updated as well. We discuss the computation complexity to complete the above updates separately.

\subsection{{Online Update of Kernel Functions}}

\begin{algorithm}[!t]
\caption{Kernel  Function  Online Update Algorithm}
\label{Update  Kernel  Functions  Online}
\KwIn{$ \Delta t$ , $\{\varepsilon^{t + \Delta t} - \varepsilon^{t}\}$ , $\{\zeta^{t+\Delta t}-\zeta^{t}\}$ , $\Phi_i(t)$'s , $\Psi_i(t)$'s , $N_i(t)$'s}
\KwOut{$\Phi_i(t+\Delta t)$'s , $\Psi_i(t+\Delta t)$'s , $N_i(t+\Delta t)$'s}
    \While{$\forall i$}{
        $\Phi_i\left( t +\Delta t \right) \leftarrow  \Phi_i\left( t  \right) \cdot exp[-\delta_0 \Delta t]$ \\
        $\tau_l \leftarrow  t$\\
        \For{$\forall \tau \in \{\varepsilon_i^{t+\Delta t}-\varepsilon_i^{t}\}$}{
            $N_i(\tau) \leftarrow  N_i(\tau_l) + 1$\\
            $\Phi_i\left( t +\Delta t \right) \leftarrow  \Phi_i\left( t + \Delta t \right) + k_0(t + \Delta t - \tau) \cdot \exp[\alpha_i N_i(\tau)]$ \\
            $ \tau_l \leftarrow  \tau$\\
        }
        $N_i(t+\Delta t) \leftarrow  N_i(\tau_l)$\\
        $\Psi_i\left( t +\Delta t \right) \leftarrow  \Psi_i\left( t  \right) \cdot exp[-\delta_1 \Delta t]$ \\
        \For{$\forall \tau' \in \{\zeta_i^{t+\Delta t}-\zeta_i^{t}\}$}{
            $\Psi_i\left( t +\Delta t \right) \leftarrow  \Psi_i\left( t + \Delta t \right) + k_1(t + \Delta t - \tau')$\\
        }
    }
\Return $\Phi_i(t+\Delta t)$'s , $\Psi_i(t+\Delta t)$'s , $N_i(t+\Delta t)$'s
\end{algorithm} 

According to Eq.~\eqref{HRS intensity function}, if there are new events, we need to update kernel functions, \emph{i.e.}, $\Phi_i(t)$ and $\Psi_i(t)$, so as to update $\lambda_i(t)$. Suppose that the time point of the last update is $t$ and the current time point to update request rates is $t+\Delta t$. Then, the computation complexity to complete the update is $O(|\varepsilon^{t+\Delta t}-\varepsilon^{t}|+|\zeta^{t+\Delta t}-\zeta^t|)$. In other words, the complexity is only a linear function with the number of incremental events.

For using exponential kernel functions, we can prove that
\begin{align}
	\begin{split}
        k_0(t+\Delta t-\tau) &= k_0(t-\tau)exp[-\delta_0\Delta t],
        \label{EQ:UpdateKernel0}
	\end{split}
	\end{align}
	\begin{align}
	\begin{split}
	k_1(t+\Delta t-\tau) &= k_1(t-\tau)exp[-\delta_1\Delta t].
	\label{EQ:UpdateKernel1}
	\end{split}
\end{align}
Note that the term $exp[-\alpha_i N_i(\tau)]$ in Eq.~\eqref{HRS intensity function} is not dependent on $t$. Thus, we can complete the update of terms $\omega_{i}  \Phi_i(t)$ and $\gamma_{i}\Psi_i(t)$ with $O(1)$ by multiplying $exp[-\delta_0\Delta t]$ and $exp[-\delta_1\Delta t]$ respectively. 
Then, we only need  to add   $|\varepsilon_i^{t+\Delta t}-\varepsilon_i^{t}|+|\zeta_i^{t+\Delta t}-\zeta_i^t|$  for each video $i$. Note that it is unnecessary to update videos without any new event. Thus, the overall computation complexity is $O(|\varepsilon^{t+\Delta t}-\varepsilon^{t}|+|\zeta^{t+\Delta t}-\zeta^t|)$.  The algorithm details for updating kernel functions are shown in Algorithm~\ref{Update  Kernel  Functions  Online}.

\begin{algorithm}[!t]
\caption{ Parameter Online Learning Algorithm}\label{Online Parameters Learning algorithm}
\KwIn{$\bm{\theta^{(0)}}$ , $T$ , $\Delta T$, $\{\varepsilon^{T+ \Delta T} - \varepsilon^{T+\frac{\ln{k_{th}}}{\delta_0}}\}$ , $\{\zeta^{T+\Delta T}-\zeta^T\}$ , $\Psi_i(T)$}
\KwOut{$\bm{\theta}^{(j)}$}
Update  $\Psi_i(t)$ where $t \in [T,T+\Delta T)$ once at first\\
$ j \leftarrow 0$  \\
\While{The termination condition is not satisfied}{
    Update $\Phi_i(t)$ ,$\Gamma_i(t)$ where $t \in [T,T+\Delta T)$ with set $\{\varepsilon^{T+\Delta T}-\varepsilon^{T+\frac{\ln{k_{th}}}{\delta_0}}\}$ \\
    $l\leftarrow 0$ \\
    $\bm{\nabla} \leftarrow [0]^{4*C}$\\
    \While{$\forall i $}{
        \For{$\Delta M$ \rm{samples}}{
            $\mathbf{t}^{(m)} \sim \text{Unif}(T,T+\Delta T)$\\
            Calculate $\hat{\lambda}_i(\mathbf{t}^{(m)})$, $\tilde{\lambda}_i(\mathbf{t}^{(m)})$ according to Eq.~(\ref{HRS intensity function}) and Eq.~(\ref{HRS intensity function positive}) with $\Phi_i(\mathbf{t}^{(m)})$ and $\Psi_i(\mathbf{t}^{(m)})$ \\
            $l\leftarrow l - \hat{\lambda}_{i}(\mathbf{t}^{(m)})$\\ \label{intensity calculate}
            \For{$\theta_i \in \{\beta_i,\omega_i,\alpha_i,\gamma_i\}$}{
                Calculate $\partial \hat{\lambda}_i(\mathbf{t}^{(m)})\,/\,\partial \theta_i$  by one of equation in Eqs.~(\ref{lambda dev betai})-(\ref{lambda dev gammai}) corresponding to $\theta_i$ \\
                $\bm{\nabla}_{\theta_i} \leftarrow \bm{\nabla}_{\theta_i} - \partial \hat{\lambda}_i(\mathbf{t}^{(m)})\,/\,\partial \theta_i$\\
            }
        }
    $l \leftarrow (\Delta T/\ \Delta M) \cdot l $  \\
    $\bm{\nabla} \leftarrow (\Delta T/\ \Delta M) \cdot  \bm{\nabla}$\\
    \For{$\forall \tau \in \{\varepsilon^{T +\Delta T}_i -\varepsilon^{T}_i\}$}{
            $l \ \leftarrow l + \log \hat{\lambda}_{i}(\tau)$\\
            \For{$\theta_i \in \{\beta_i,\omega_i,\alpha_i,\gamma_i\}$}{
                $\bm{\nabla}_{\theta_i} \leftarrow \bm{\nabla}_{\theta_i} + (\partial \hat{\lambda}_i(\tau)\,/\,\partial \theta_i) / \ \hat{\lambda}_{i}(\tau)$\\
            }
         } 
    }
    Update $\bm{\theta}^{(j+1)} \leftarrow \bm{\theta}^{(j)}$ by adopting L-BFGS algorithm with $l$, $\bm{\nabla}$ and the penalty term \\
    $j \leftarrow j+1$
}
$T \leftarrow T + \Delta T$\;
\Return $\bm{\theta}^{(j)}$
\end{algorithm}
\subsection{Online Update of Parameters}

To update parameters in accordance with new events, it is necessary to update gradients based on Eq.~\eqref{lambda dev betai}-\eqref{lambda dev gammai} and execute the GD algorithm again to obtain updated parameters. To avoid confusing with the kernel function update,  we suppose the time point of the last parameter update is $T$ and the current time point is $T+\Delta T$\footnote{Note that the update frequency of parameters is different from that of video request rates.}.

In  Eqs.~\eqref{lambda dev betai}-\eqref{lambda dev gammai}, we can also see kernel functions $\Phi_i(t)$'s, $\Psi_i(t)$'s and $\Gamma_i(t)$'s. Therefore, the update of them will be firstly introduced.
 If maintaining a fixed value during the learning process, the kernel functions $\Psi_i(t)$'s can be simply updated by incremental new events based on the discussion in the last subsection. 
 However, as for $\Phi_i(t)$'s and $\Gamma_i(t)$'s, we need to scan all historical events again to compute their values when the parameter $\alpha$ has been updated, which results in very high computation complexity.  
 To reduce the additional complexity, we propose to use a threshold $k_{th}$ to truncate the sum of kernel functions. $k_{th}$ is a very small number. If $k_0(t-\tau)<k_{th}$ where $t$ is the current time and $\tau$ is the occurrence of a particular event, it implies that the influence of the historical event before time $\tau$ is negligible. 
 Therefore, it is trivial to ignore this event so that the computation complexity will not continuously grow with time. 
 Given $k_{th}$ and the current update time $T+\Delta T$, it is easy to verify that events between time $[T+ \frac{\ln{k_{th}}}{\delta_0},T+\Delta T)$ needs to be involved to update $\Phi_i(t)$'s and $\Gamma_i(t)$'s.
 Thus, the upper bound of computation complexity for kernel renewal is $O(|\varepsilon^{T+\Delta T}-\varepsilon^{T + \frac{\ln{k_{th}}}{\delta_0}}|+|\zeta^{T+\Delta T}-\zeta^{T}|)$.

Next, we introduce the update of all gradients in Eq.~\eqref{Eq:l dev theta_i}. The first term relevant to recent new events can be updated in $O(|\varepsilon^{T+\Delta T} -\varepsilon^{T}|)$. Furthermore, the Monte Carlo sampling term needs to be trimmed during online learning process, \emph{i.e.}, $\frac{\Delta T }{\Delta M}\sum_{m=1}^{\Delta M} \frac{\partial \hat{\lambda}_i(\mathbf{t}^{(m)})}{\partial \theta_i}$. We suppose there are $\Delta M$ new samples during the period from $[T, T+\Delta T)$, where $\Delta M/M = \Delta T/T$ since we update the parameters in a shorter time window. Therefore, we can conclude that the complexity to calculate  the Monte Carlo sampling term is $O(C \cdot \Delta M)$ with  updated kernels. Here $C$ is the total number of videos in the system.

By wrapping up our analysis, the overall computation complexity to update all gradients is $O(|\zeta^{T+\Delta T}-\zeta^T|+|\varepsilon^{T+\Delta T}-\varepsilon^{T+\frac{\ln{k_{th}}}{\delta_0}}|+|\varepsilon^{T+\Delta T} -\varepsilon^{T}|+C \cdot \Delta M)$ in each iteration.   
Note that this is also an upper bound of the computation complexity. 
In the end, we present the detailed online learning algorithm for training the HRS model in Algorithm \ref{Online Parameters Learning algorithm}.

\subsection{{Framework of HRS Edge Caching System}} 
	\begin{figure}[!t]
    \centering
    \includegraphics[width=3.5in]{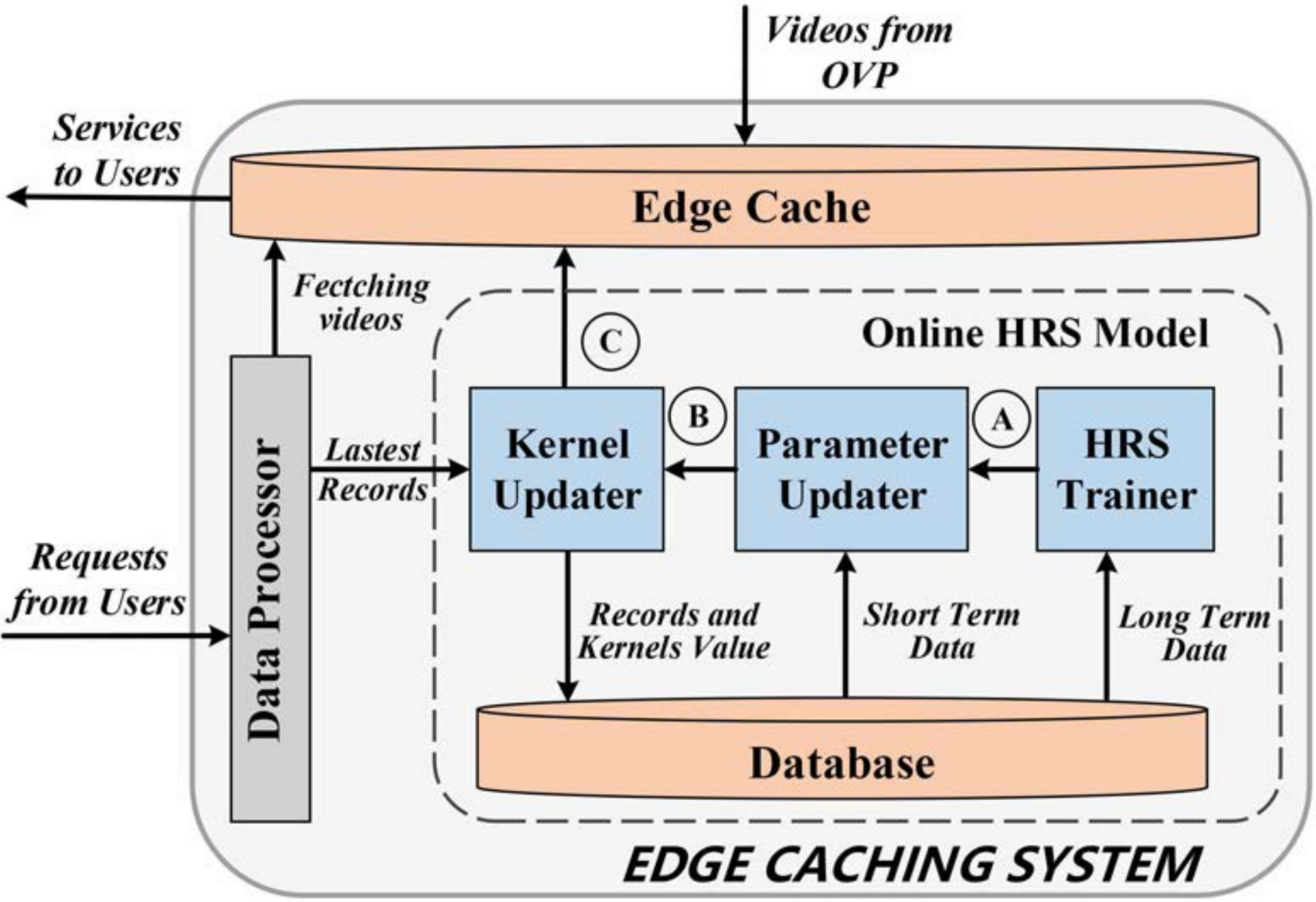}
    \caption{Modules of an edge cache system implementing online HRS algorithm. A representative workflow is presented in order.}
    \label{HRSMODLE}
    \end{figure}
    
In Fig.~\ref{HRSMODLE}, we describe the framework of our system including an \textit{Online HRS model}, a \textit{Data Processor} and an \textit{Edge Cache}.

As shown in Fig.~\ref{HRSMODLE},  
the \textit{Data Processor}  is responsible for preprocessing the request records. 
It is also responsible for recording positive and negative events.
The \textit{Online HRS model} utilizes the user request records from the \textit{Data Processor} to periodically update kernel functions and the HRS model parameters. 
the \textit{Edge Cache} can periodically update the cached videos based on the update of prediction results generated by the HRS model. 
Based on the updated request rates of all videos, the \textit{Edge Cache} maintains videos with the highest request rates in its cache until the cache is fully occupied. 
To reduce the computation overhead, the \textit{Edge Cache} can only  re-rank videos with new request rates updated by the \textit{Online HRS model}. 

 The \textit{Online HRS model} can be further decomposed into three parts, including \textit{HRS Trainer}, \textit{Parameter Updater} and \textit{Kernel Updater}.  A typically workflow through these three parts is shown as follows:
\begin{itemize}
       \item[A)] This is the first step to deploy the HRS algorithm on an edge node.  The HRS model can be initially trained by the \textit{HRS Trainer} with the long term request records according to the method introduced in Section \ref{offline training method}. 
       \item[B)] With the arrival of new video requests, the parameters such as $\alpha_i$ and $\beta_i$ in the HRS model can be refined by the  \textit{Parameter Updater} according to Alg.\ref{Online Parameters Learning algorithm}. Thus, these parameters reflect not only long-term  but also short-term popularity trends of videos. To avoid overfitting, these parameters should not be updated over frequently. In our experiments, we update  parameters for every a few number of days.
       \item[C)] The user view interest may change very quickly with time, which can be captured by the renewal of prediction on video request rates through frequently updating the kernel functions. 
       With the incremental user requests, the \textit{Kernel Updater} can efficiently update the kernel functions to trace the latest user request interest. 
\end{itemize} 


\section{Evaluation\label{Evaluation}}

We evaluate the performance of our HRS algorithm by conducting  experiments with real traces collected from the Tencent Video.

\subsection{{Dataset}}
Tencent video \footnote{Tencent Video: https://v.qq.com} is one of the largest online video streaming platforms in China.  We collected a total of 30 days of request records from  Nov 01, 2014 to Nov 30, 2014. 
After data cleaning, encoding and masking, we randomly sample a dataset which contains a population $C = 20K$ of unique videos from 5 cities in Guangdong province in China. There is a  total  number of $K = 15.84M(15,841,209)$ request records in this dataset and we make the dataset publicly available on GitHub\footnote{https://github.com/zhangxzh9/HRS\_Datasets}.
Each request record in our dataset is represented by the metadata {\small $\left \langle \;VideoID, UserID, TIME, PROVINCE, CITY\;\right \rangle$}. Given the lack of negative events, the set of negative events is generated from request records as follows: if a video stays cold without being requested for a period, a negative event of this video will be marked in the dataset. Empirically, the period is set as 12 hours.


We divided the dataset into five parts based on the date of request records for cross-validation and hyper-parameters selection following the Forward-Chaining trick\cite{Bergmeir2012}.
Each part includes the request records in six days.
In the first fold, Part \uppercase\expandafter{\romannumeral1}, including the request traces in the first 6 days, is used as the training set; while Part \uppercase\expandafter{\romannumeral2}, with the records from day 6 to 12, is used as validation set.  Part \uppercase\expandafter{\romannumeral3}, including the request records of the next 6 days, is used as the test set. 
In the second fold, Part \uppercase\expandafter{\romannumeral1} and \uppercase\expandafter{\romannumeral2} together serve as the training set, while Part \uppercase\expandafter{\romannumeral3} and \uppercase\expandafter{\romannumeral4} are used as the validation set and test set respectively. 
Finally, we employ Parts \uppercase\expandafter{\romannumeral1}-\uppercase\expandafter{\romannumeral3} as the training set and the rest two parts as the validation and test set respectively in the third fold.


\subsection{{Evaluation Metrics}}
We employ two metrics for evaluation. 
\begin{itemize}
\item \textbf{Cache Hit Rate} is defined as the number of requests hit by the videos cached on the edge server divided by the total number of requests issued by users. If the HRS runs independently on multiple edge servers, the overall cache hit rate of the whole system will be calculated by the weighted average hit rates of multiple edge servers.

\item \textbf{Execution Time}: 
Due to the possibility that the edge server is with very limited computing resource, it is desirable that the computation load of the  caching algorithm  is under control so that cached videos can be updated timely.  Thus, we use  the execution time of each algorithm as the second evaluation metric. 

\end{itemize}

\subsection{{Baselines}}
We compare the performance of HRS with the following baselines:
\begin{itemize}
    \item \textbf{LRU (Least Recently Used)}, which always replaces the video that was requested least recently with the newly requested one when the cache is full.
    \item \textbf{OPLFU (Optimal Predictive Least Frequently Used)}\cite{Famaey2013}, which is a variant of LFU. Different from LFU, it predicts the future popularity by matching and using one of Linear, Power-Law, Exponential and Gaussian functions. Caching server maintains the cache list based on the estimated future popularity determined by the selected function. Due to the high computation complexity, we only use Linear, Power-Law and Exponential functions in our experiments.
    \item \textbf{POC (Popcaching)} \cite{Li2016a}, which learns the relationship between the popularity of videos and the context features of requests, and stores all features in the Learning Database for video popularity prediction. Once a request arrives, POC will update the features of the requested video online and predicts video popularity by searching the Learning Database with the context features. we set the number of requests in the past 1 hour, 6 hours, 1 day as the first three features while 10 days, 15 days and 20 days as the fourth dimension feature for three folds, respectively.
    
    \item \textbf{LHD (Least Hit Density)} \cite{Beckmann2018a}, which is a rather rigorous eviction policy to determine which video should be cached. LHD predicts potential hits-per-space-consumed (hit density) for all videos using conditional probability and eliminates  videos with the least contribution to the cache hit rate. An public implementation of the LHD algorithm in GitHub\footnote{https://github.com/CMU-CORGI/LHD} can be obtained and we reuse it in our experiments by setting all parameters as default values.
    \item \textbf{DPC (DeepCache)} \cite{Narayanan2018}, which predicts video popularity trend by leveraging the LSTM Encoder-Decoder model. An $M$-length input sequence with $d$-dimensional feature vector representing the popularity of $d$ videos is required for the model.  A $K$-length  sequence will be exported for prediction. Here $M$ and $K$ are hyper-parameters for model. All model settings are the same as the work \cite{Narayanan2018} in our experiments.
    
    \item \textbf{Optimal (Optimal Caching)}, which is achieved by assuming that  all future requests are known so that the edge server can always make the optimal caching decisions. It is not a realistic algorithm, but can be used to evaluate the improvement space of each caching algorithm. 
\end{itemize}

\subsection{{Experimental Settings}}
We simulate a video caching system shown in Fig.~(\ref{HRSMODLE}) to evaluate all caching algorithms. In the gradient descent algorithm, there are six hyper-parameters which are initialized as $\delta_0=0.5$, $\delta_1 = 1.5 $ and  $\rho_{\beta}=\rho_{\alpha}=\rho_{\omega} =\rho_{\gamma}=e^5$ empirically. Their values will be determined through validation.
For all parameters $\boldsymbol{\theta}$ in HRS, their initial values are set as $1$, except that 
the initial values of $\boldsymbol{\gamma}$ are set as 0.1, referring to the settings in some previous papers\cite{Rizoiu2017,Ertekin2015,Xu2017}.
Moreover, the number of sample times in the Monte Carlo estimation is set as $144,000$ times for every day, \emph{e.g.}, we set $M =1,728,000$ in the first fold, and set $M =2,592,000$ and $M =3,456,000$ for the second and third folds respectively. The time interval $\Delta T$ to update the online HRS model is set as two days.
{The iteration process in the gradient descent algorithm will be terminated if the improvement of cache hit rate in the validation set is negligible after an iteration. }

By default, the time interval to update kernel functions is set as  1 hour, and the truncating threshold $k_{th}$ is set as $e^{-9}$ for parameter online learning. Some detailed experiments are conducted to study the influence of these two parameters.

Furthermore, all algorithms except LHD are programmed with Python\cite{PythonSoftwareFoundation2016} and executed in Jupyter Notebook \cite{ProjectJupyter2019} with a single process. As for LHD, we reuse the code and estimate the execution time according to \cite{Beckmann2018a}.
Besides, the execution time is measured on an Intel server with Xeon(R) CPU E5-2670 @ 2.60GHz. 

\subsection{{Experimental Results}}

\subsubsection{Comparison of Cache Hit Rate}
\begin{figure*} 
    \centering
	  \subfloat[Cache hit rate with small cache at city level.]{\label{city_small hitrate}
       \includegraphics[width=0.48\linewidth]{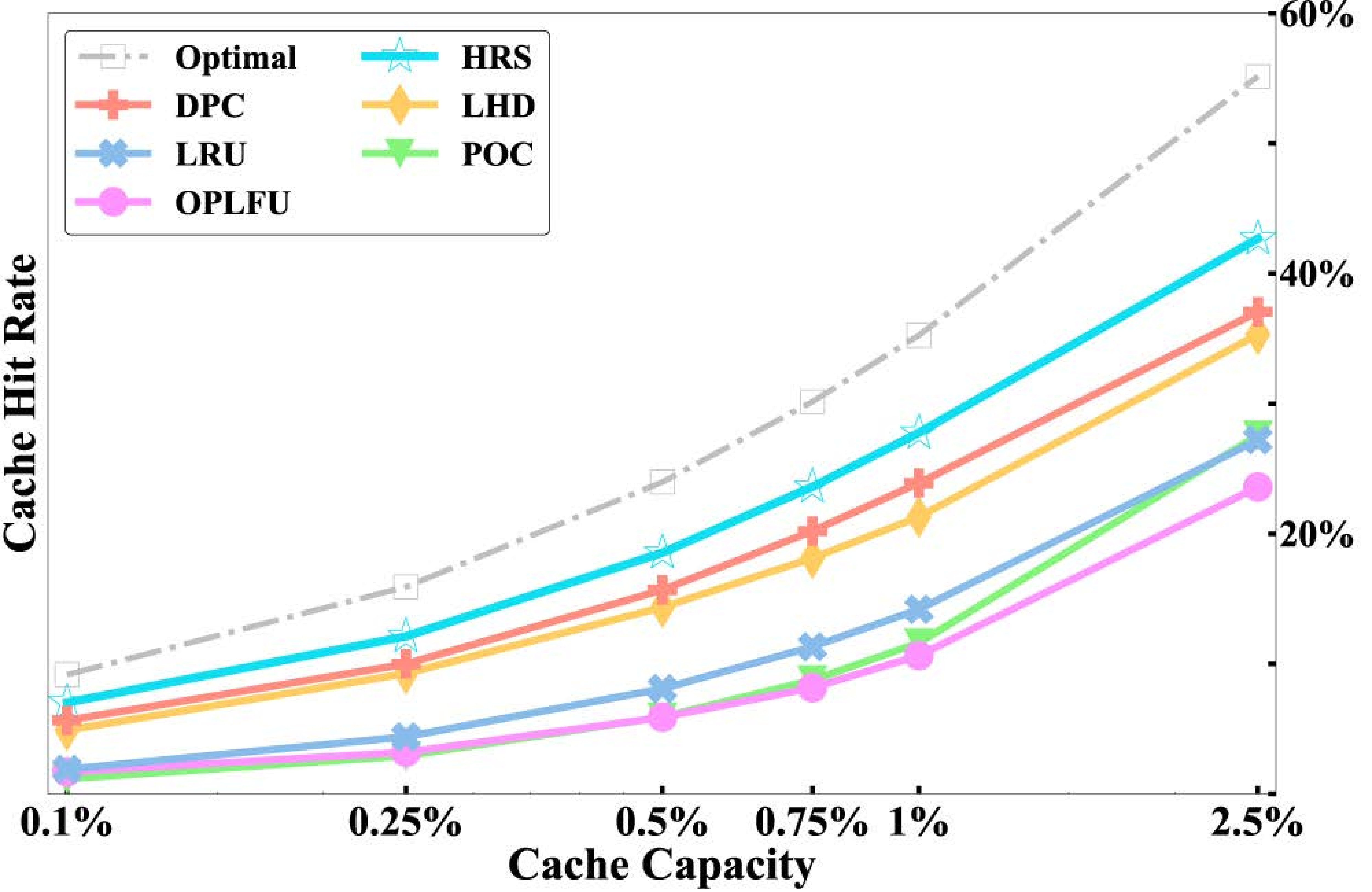}
    }\hfill
	  \subfloat[Cache hit rate with large cache at city level.]{\label{city_large hitrate}
    \includegraphics[width=0.48\linewidth]{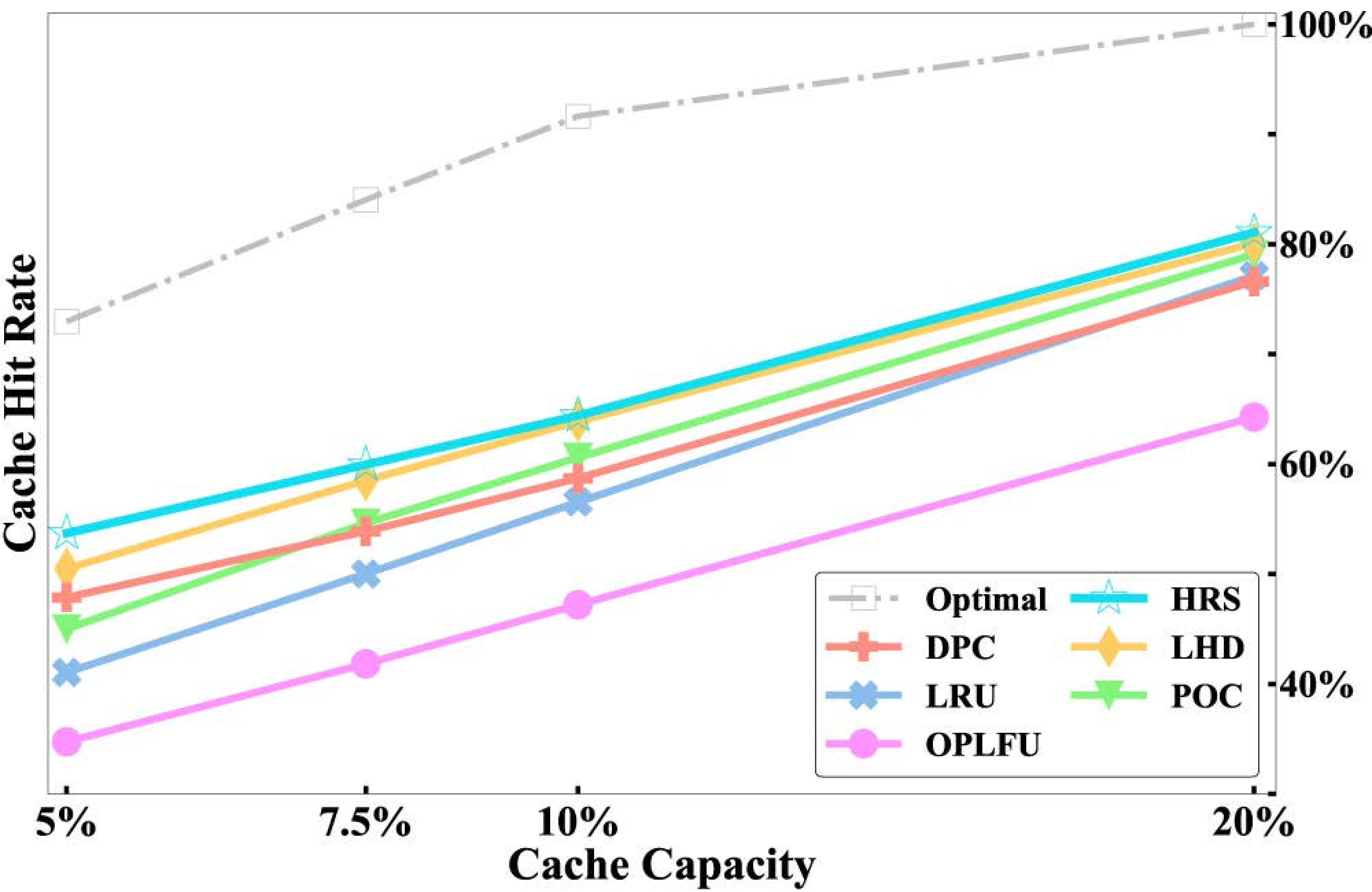}
    }
	  \caption{Cache hit rate for HRS and other baseline algorithms when varying the cache capacity (size)  from $S=0.1\% \ (20)$  to $S=25\% \ (5000)$ at the city level. For clarity, Fig.~(\ref{city hitrate}a) and Fig.~(\ref{city hitrate}b) show the performance under the small and large capacity respectively.}
	  \label{city hitrate} 
\end{figure*}

\begin{figure*} 
    \centering
	  \subfloat[Cache hit rate with small cache at province level.]{\label{prov_small hitrate}
       \includegraphics[width=0.48\linewidth]{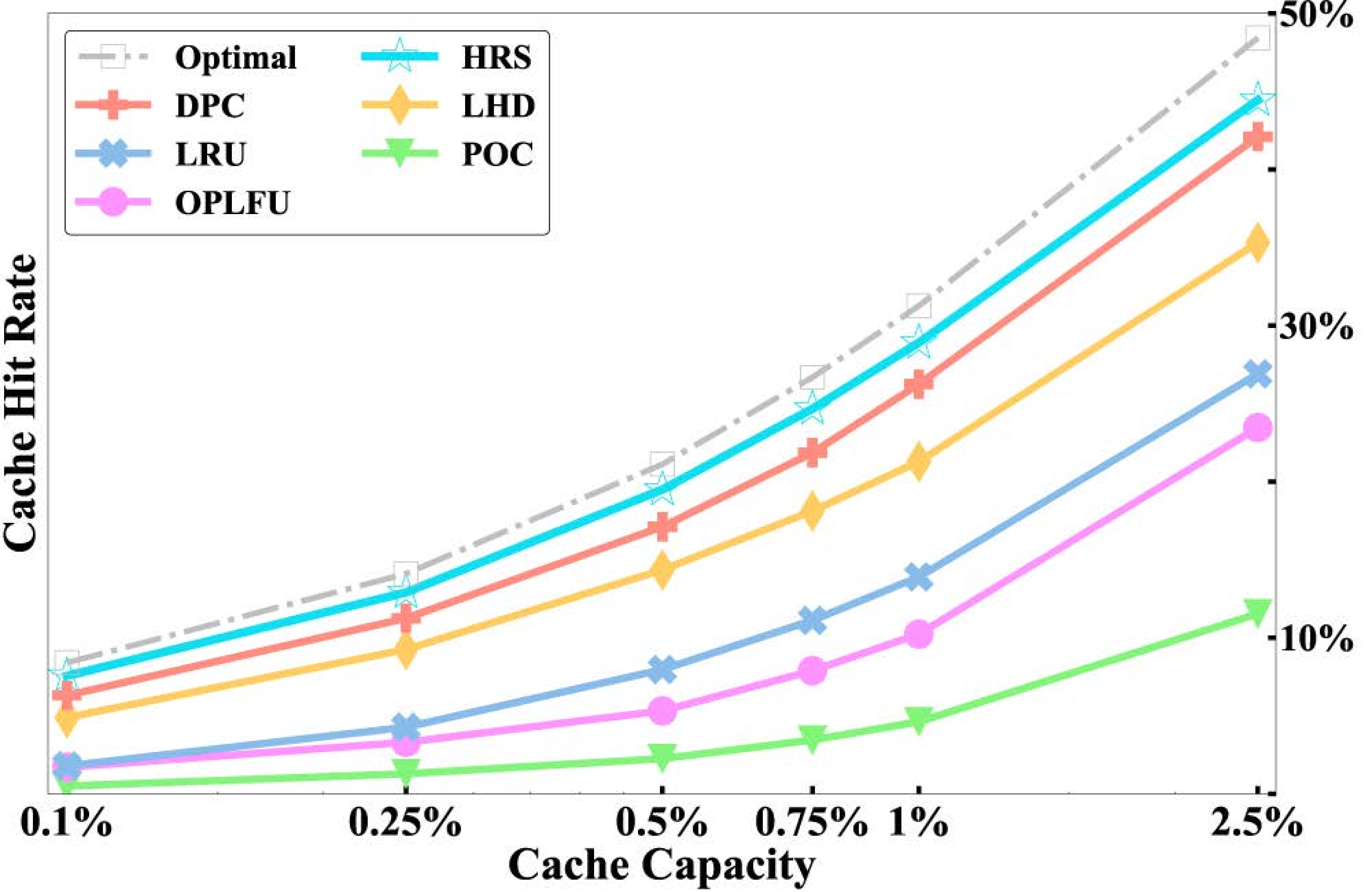}
    }\hfill
	  \subfloat[Cache hit rate with large cache at province level.]{\label{prov_large hitrate}
        \includegraphics[width=0.48\linewidth]{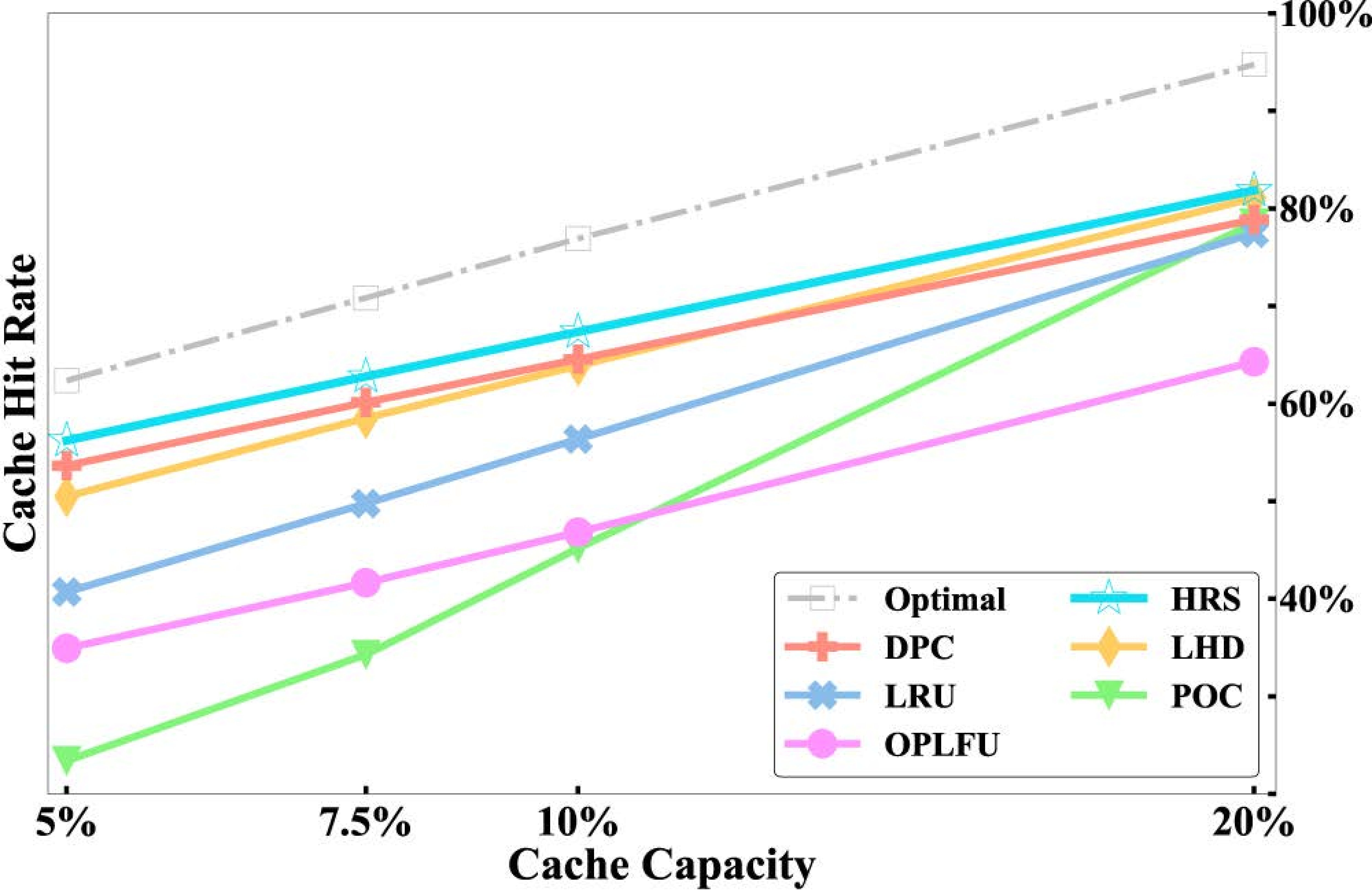}
    }
	  \caption{Cache hit rate for HRS and other baseline algorithms when varying the cache capacity (size) from $S=0.1\%\ (20)$  to $S=25\% \ (5000)$ at the province level.  For clarity, Fig.~(\ref{prov hitrate}a) and Fig.~(\ref{prov hitrate}b) show the performance under the small and large capacity respectively.}
	  \label{prov hitrate} 
\end{figure*}

We first evaluate the HRS algorithm with other baseline caching algorithms through experiments by varying the caching capacity from 0.1\% to 25\% of the total number of videos (\emph{i.e.}, the caching size is varied from 20 to 5K videos). The experiment results of 5 cities are presented in Fig.~\ref{city hitrate} with the y-axis representing the averaged cache hit rate and {Fig.~\ref{prov hitrate} shows the results of cache hit rate of a single server at the province level.}

Through the comparison, we can see that:
\begin{itemize}
    \item HRS outperforms all other baseline algorithms evaluated in terms of the cache hit rate under all cache sizes over the test time, with an overall average of 12.3\% improvement. DPC is the second best one in most cases.
    \item  It is an effective approach to reduce the Internet traffic by caching videos with HRS. For example, the cache hit rate is more than 45\% by only caching 2.5\% video contents implying that the Internet traffic can be reduced by 45\%.
    \item It is more efficient to utilize the caching capacity by using  HRS. To demonstrate this point, let us see a specific case with a target cache hit rate of 10\% in Fig.~\ref{city hitrate}. In this case, HRS requires a cache size of 40 videos. In comparison,  DPC/LHD needs nearly 2-2.4 times cache capacity to achieve the same goal. The performance improvement against the second best solution exceeds  124\% with 0.1\% limited caching capacity, showing the outstanding ability of HRS to predict the most popular video when the resource is constrained and a more accurate decision is needed.
    \item  Compared to other baseline algorithms at the province level, HRS also achieves an overall average of 8.4\% improvement. The HRS model performs better at the city level than at the province level for the reason that video popularity trends are more accurately  reflected by leveraging the SC term in city edge servers. In fact, the HRS model performs better if the request rate is higher. 
\end{itemize}

\begin{figure}[!t]
\centering
\includegraphics[width=3.5in]{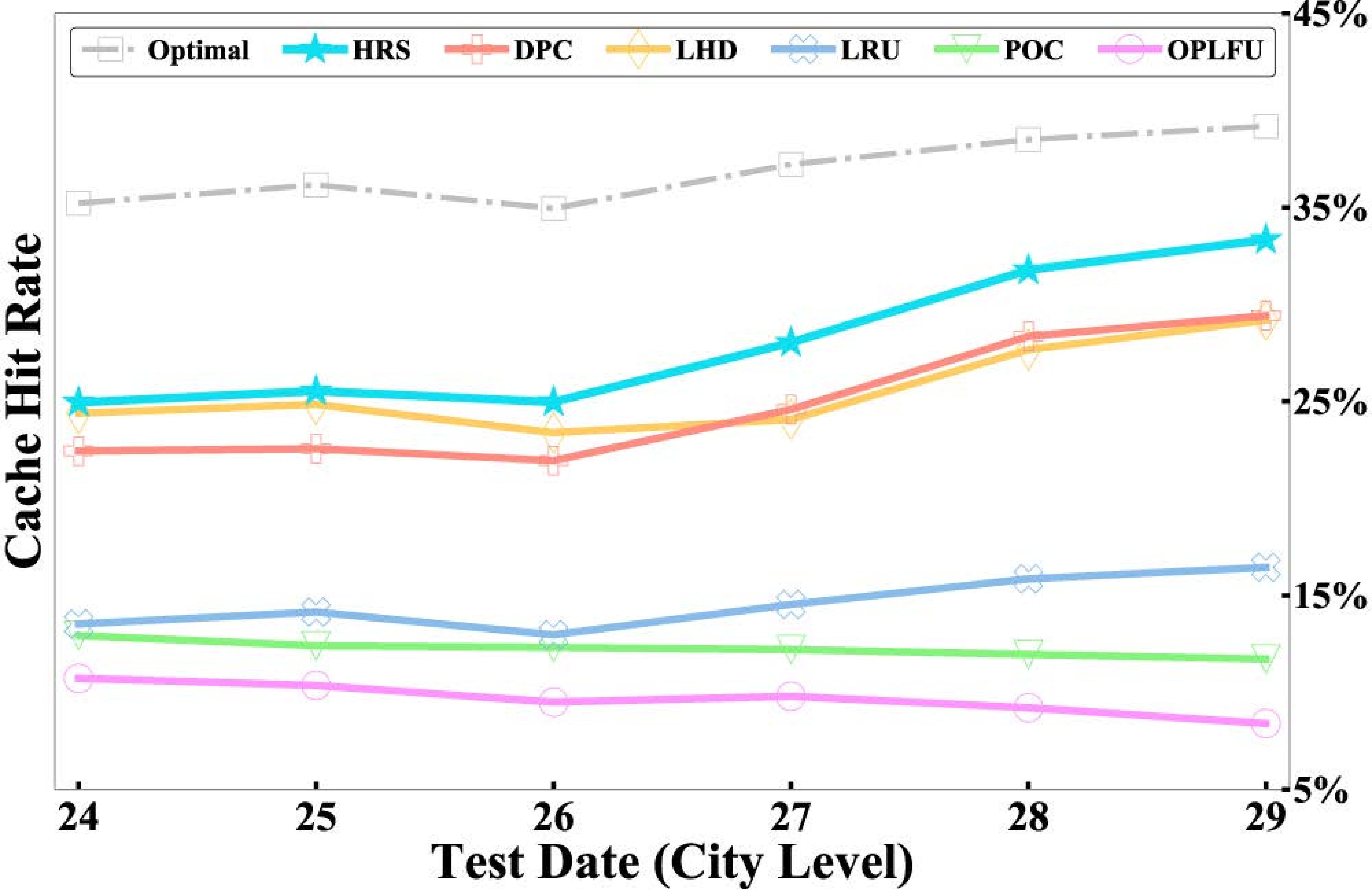}
\caption{Average cache hit rate of each day over the test period. Each point in the figure shows the average cache hit rate over a day. The cache capacity is fixed at $S=1\%\ (200)$.}
\label{day hit rate}
\end{figure}

Moreover, to check the stability of each video caching algorithm, we plot the cache hit rate over time  with a fixed caching capacity $S=200$ (equivalent to about 1\% of total videos). The results are presented in  Fig.~\ref{day hit rate}, showing the averaged cache hit rate of all algorithms versus the date. In other words, each point  in the figure represents the average cache hit rate over a day.
From the results shown in Fig.~\ref{day hit rate}, we can draw a conclusion that HRS is always the best one achieving the highest cache  hit rate among video caching algorithms except the Optimal one indicating the gain of HRS is very stable over time. 

\subsubsection{Setting  of Hyper-parameters}

We study the sensitivity of two crucial parameters: $\Delta t$ and $k_{th}$ in Tables \ref{update frequency} and \ref{truncated threshold} to see how these two hyper-parameters affect the video caching performance. All other hyper-parameters are kept unchanged as we vary $\Delta t$ and $k_{th}$.

We repeat the experiments presented in Fig.~\ref{city hitrate} by setting different values for $\Delta t$. 
The parameter $\Delta t$ indicates how frequently the HRS model updates the kernel functions. 
As we can see in Table \ref{update frequency}, the cache hit rate is higher if $\Delta t$ is smaller because the latest user trends can be captured in time with a smaller $\Delta t$. It also confirms that the user interest is very dynamic over time. 
However, it is more reasonable to  set $\Delta t = 1$  hour since the improvement using $\Delta t = 0.5$ hour is small but with higher time complexity . 

\begin{table}[!bp]
\renewcommand{\arraystretch}{2}
\setlength\tabcolsep{3pt}
\caption{Cache hit rate (\%) under different $\Delta t$ (hour(s)). The setting is the same as that in Fig.~\ref{city hitrate} except $\Delta t$.}
\begin{center}
\begin{tabular}{|p{1.25cm}<{\centering}||c|c|c|c|c|c|c|}
\hline
\small{\textbf{Update}}&\multicolumn{7}{c|}{\normalsize{\textbf{Cache Capacity}}} \\
\cline{2-8} 
\small{\textbf{Interval}}   &\makecell[c]{20\\(0.1\%)}&\makecell[c]{50\\(0.25\%)}&\makecell[c]{100\\(0.5\%)}&\makecell[c]{200\\(1\%)}&\makecell[c]{500\\(2.5\%)}&\makecell[c]{1000\\(5\%)}&\makecell[c]{5000\\(25\%)}\\
\hline
$\Delta t=4.0$&6.159 &10.934&17.120&26.403&41.415&52.750&79.149\\
\hline
$\Delta t=1.0$&\textbf{7.015}&\textbf{12.110}&\textbf{18.538}&\textbf{27.754}&\textbf{42.693}&53.712&80.112\\
\hline
$\Delta t=0.5$&6.988&12.084&18.442&27.607&42.577&\textbf{54.183}&\textbf{80.700}\\
\hline
\end{tabular}
\label{update frequency}
\end{center}
\end{table}


In Table~\ref{truncated threshold}, we further investigate the influence of $k_{th}$ on the cache hit rate. $k_{th}$ is negligible and can be discarded to control the computation complexity. We reuse the setting of the experiment in Fig.~\ref{city hitrate} except varying  $k_{th}$. As we can see from Table~\ref{truncated threshold}, the overall cache hit rate is better if $k_{th}$ is smaller since more kernel functions are reserved for computation. Because the cache hit rate is very close by setting $k_{th}$ equal to $e^{-9}$ or $e^{-10}$, we finally set $k_{th}=e^{-9}$ in our experiments with lower time complexity. 


\subsubsection{Comparison of Execution Time}

\begin{figure}[!t]
\centering
\includegraphics[width=3.5in]{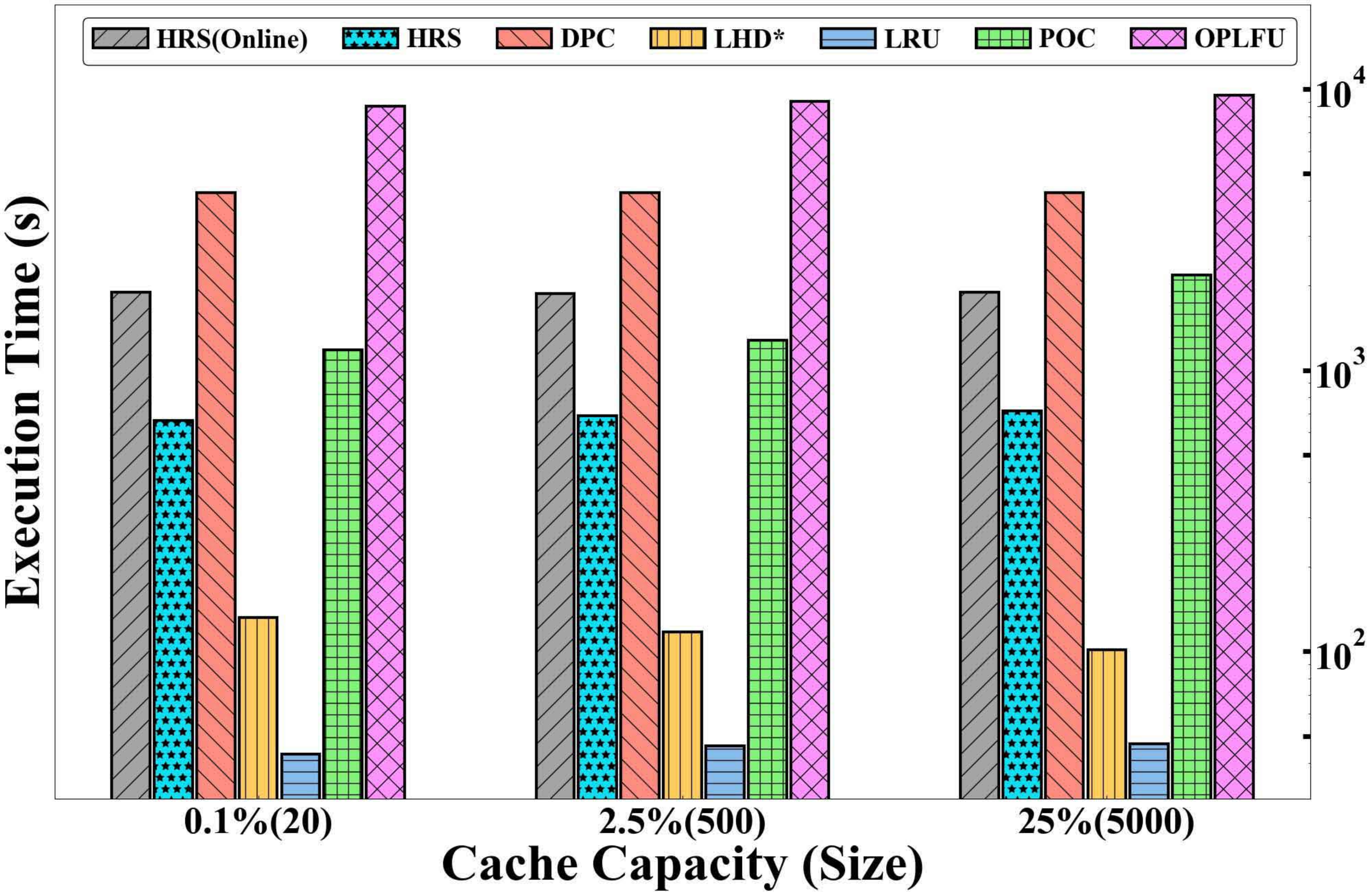}
\caption{Comparison of execution time under different cache sizes. Results show the total time(s) for 5 cities to maintain the caching list during the test period.}
\label{execution time}
\end{figure}

We conduct experiments to evaluate the execution time of each video caching algorithm under various caching sizes in Fig.~\ref{execution time}. Notably, HRS(Online) and HRS are conducted to examine the   influence of online algorithm on computation complexity. 
Eventually, the tests are carried out on 5 cities and all test periods in the three folds are considered to achieve convincing results.


As we see from Fig.~\ref{execution time}, the heuristically designed algorithms, \emph{i.e.}, LHD and LRU,  achieve the lowest execution time. However, HRS is the best one compared with other proactive video caching algorithms, \emph{i.e.}, DPC, POC and OPLFU. The execution time of the online HRS-based algorithm is very short since the truncated threshold for kernel renewal can control the computation complexity. 
 Moreover, this experiment results indicate the feasibility of HRS for online video caching. 

\begin{table}[!bp]
\renewcommand{\arraystretch}{2}
\setlength\tabcolsep{2.5pt}
\caption{Cache hit rate(\%) under different values for the threshold $k_{th}$. The setting is the same as that in Fig.~\ref{city hitrate} expect $k_{th}$.}
\begin{center}
\begin{tabular}{|p{1.45cm}<{\centering}||c|c|c|c|c|c|c|}
\hline
\textbf{Truncating}&\multicolumn{7}{c|}{\normalsize{\textbf{Cache Capacity}}} \\
\cline{2-8} 
\textbf{Threshold}&\makecell[c]{20\\(0.1\%)}&\makecell[c]{50\\(0.25\%)}&\makecell[c]{100\\(0.5\%)}&\makecell[c]{200\\(1\%)}&\makecell[c]{500\\(2.5\%)}&\makecell[c]{1000\\(5\%)}&\makecell[c]{5000\\(25\%)}\\
\hline
$k_{th} = e^{-8}$&6.967&12.022&18.344&27.405&42.119&53.049&79.961\\
\hline
$k_{th} = e^{-9}$&7.015&12.110&18.538&27.754&\textbf{42.693}&\textbf{53.712}&80.112\\
\hline
$k_{th} = e^{-10}$&\textbf{7.025}&\textbf{12.126}&\textbf{18.570}&\textbf{27.777}&42.686&53.631&\textbf{80.148}\\
\hline
\end{tabular}
\label{truncated threshold}
\end{center}
\end{table}

\section{Related Work\label{Related Work}}

\subsection{Video Caching}
Caching at the edge is an effective way to alleviate the backhaul burden and reduce the response time. In recent years, more and more research interest has been devoted to investigating the caching problem on edge servers. 
Golrezaei \emph{et al.} presented a novel architecture of distributed caching with the D2D collaboration to improve the throughput of wireless networks \cite{Golrezaei2013}. 
Gregori \emph{et al.} executed the caching strategies on small base stations or user terminals by D2D communications \cite{Gregori2016}. 
Different from caching by D2D communication, caching at the edge has more potential to make precise decisions with features of edge servers. 
Poularakis \emph{et al.} formulated a joint routing and caching problem guaranteed by an approximation algorithm to improve the percentage of requests served by small base stations (SBSs)\cite{Poularakis2014}. 
Jiang \emph{et al.} developed a cooperative cache and delivery strategy in heterogeneous 5G mobile communication networks \cite{Jiang2017}. 
Yang \emph{et al.} devised an online algorithm which estimates  future popularity by location customized caching schemes in mobile edge servers \cite{Yang2019}.  
Moreover, with the ability to learn the context-specific content popularity online, a context-aware proactive caching algorithm in wireless networks was introduced by Muller \emph{et al.} \cite{Muller2017}.

With the explosive growth of video population, it is urgent to develop more intelligent video caching algorithms by identifying popularity patterns in historical records. 
It was summarized in \cite{Sitaraman2014} and \cite{Goian2019} that diverse approaches for content caching have been implemented in the Internet nowadays. However, less attention has been allocated to optimize the caching methods and most of them were deployed based on heuristic algorithms such as LRU, LFU and their variants \cite{Jaleel2010,Famaey2013, Shafiq2014}, which are lightweight but inaccurate, and thus often fail to seize viewers’ diverse and highly dynamic interests. 

Some proactive models including regression models \cite{Wang2012}, auto regressive integrated moving average \cite{Niu2011} and classification models \cite{Rowe2011} were proposed to forecast the popularity of content. 
{Moreover, quite a few learning-driven caching algorithms were proposed for some special application scenarios.} 
Wu \emph{et al.} proposed an optimization-based approach with the aim to balance the cache hit rate and cache replacement cost\cite{Wu2015}.
Wang \emph{et al.} developed a transfer learning algorithm to model the prominence of video content from social streams \cite{Roy2013a}, while  Roy \emph{et al.} proposed a novel context-aware popularity prediction policy based on federated learning\cite{Wu2020}.

Besides, with the rapid development of deep learning, a significant amount of research efforts has been devoted to predicting content popularity  using the neural network model.
Tanzil \emph{et al.} adopted a neural network model to estimate the popularity of contents and select the physical cache size as well as the place for storing contents\cite{Tanzil2017}. 
Feng \emph{et al.} proposed a simplified Bi-LSTM  (bidirectional long short-term memory) neural network to predict the corresponding popularity profile for every content class\cite{Feng2019}. 
LSTM was also implemented for content caching in \cite{Narayanan2018}.
However, NN-based models typically require a large number of historical records for tuning the extensive parameters. But with the sparse requested records of cold videos, it is not easy to learn an appropriate model for prediction.
Further, because the popularity distribution of contents may constantly change over time\cite{Dhar2011}, it is difficult to make decisions based on outdated dataset.  
Thus, some online learning models which are more responsive to continuously changing trends of content popularity were proposed in\cite{Li2016a,Muller2017,Yang2019}.

\subsection{{Point Process}}
The point processes are frequently used to model a series of superficially random events in order to reveal the underlying trends or predict future events. 
Bharath \emph{et al.} considered a learning-driven approach with independent Poisson point processes in a heterogenous caching architecture \cite{Bharath2016}.
Shang \emph{et al.}\cite{Shang2018} formulated a model to obtain a large-scale user-item interactions by utilizing point process models. 
Xu \emph{et al.} \cite{Xu2017} modeled user-item interactions via superposed Hawkes processes, a  kind of classic point process model, to improve recommendation performance. 
More applications of point processes in recommendation systems can be found in \cite{Du2015, Dai2016}. 
Furthermore, point processes have been applied to study social networks between individual users and their neighbors in \cite{Zhou2013}. 
Ertekin \emph{et al.} \cite{Ertekin2015} used reactive point process  to predict power-grid failures, and provide a benefit-and-cost analysis for different proactive maintenance schemes. 
Mei \emph{et al.} \cite{Mei2017a} proposed a novel model which was a combination of point processes and neural networks to improve prediction accuracy for future events. 
The reason why point processes have been applied in predicting discrete events in the future lies in that the occurrence of a past event often gives a temporary boost to the occurrence probability of events of the same type in the future. Naturally, the video request records can be regarded as time series events, which can be modeled by point processes. However, there is very limited work that explored the utilization of point process models to improve video caching decisions, which is the  motivation of our work.

\section{Conclusion}
In this work, we propose a novel HRS model to make  video caching decisions for edge servers in online video systems.  HRS is developed by combing the Hawkes process, reactive process and self-correcting process to model the future request rate of a video based on the historical request events. The HRS model parameters can be determined through maximizing the Log-likelihood of past events, and detailed iterative algorithms are provided. In view of the dynamics of user requests, an online HRS-based algorithm is further proposed, which can process the request events in an incremental manner. In the end, we conduct extensive experiments through real video traces collected from Tencent Video to evaluate the performance of HRS. In comparison with other baselines, HRS not only achieves the highest cache hit rate, but also maintains low computation overhead.



%
%

\ifCLASSOPTIONcaptionsoff
  \newpage
\fi

\bibliographystyle{IEEEtran}
\bibliography{IEEEabrv,ref}

\begin{IEEEbiography}[{\includegraphics[width=1in,height=1.25in,clip,keepaspectratio]{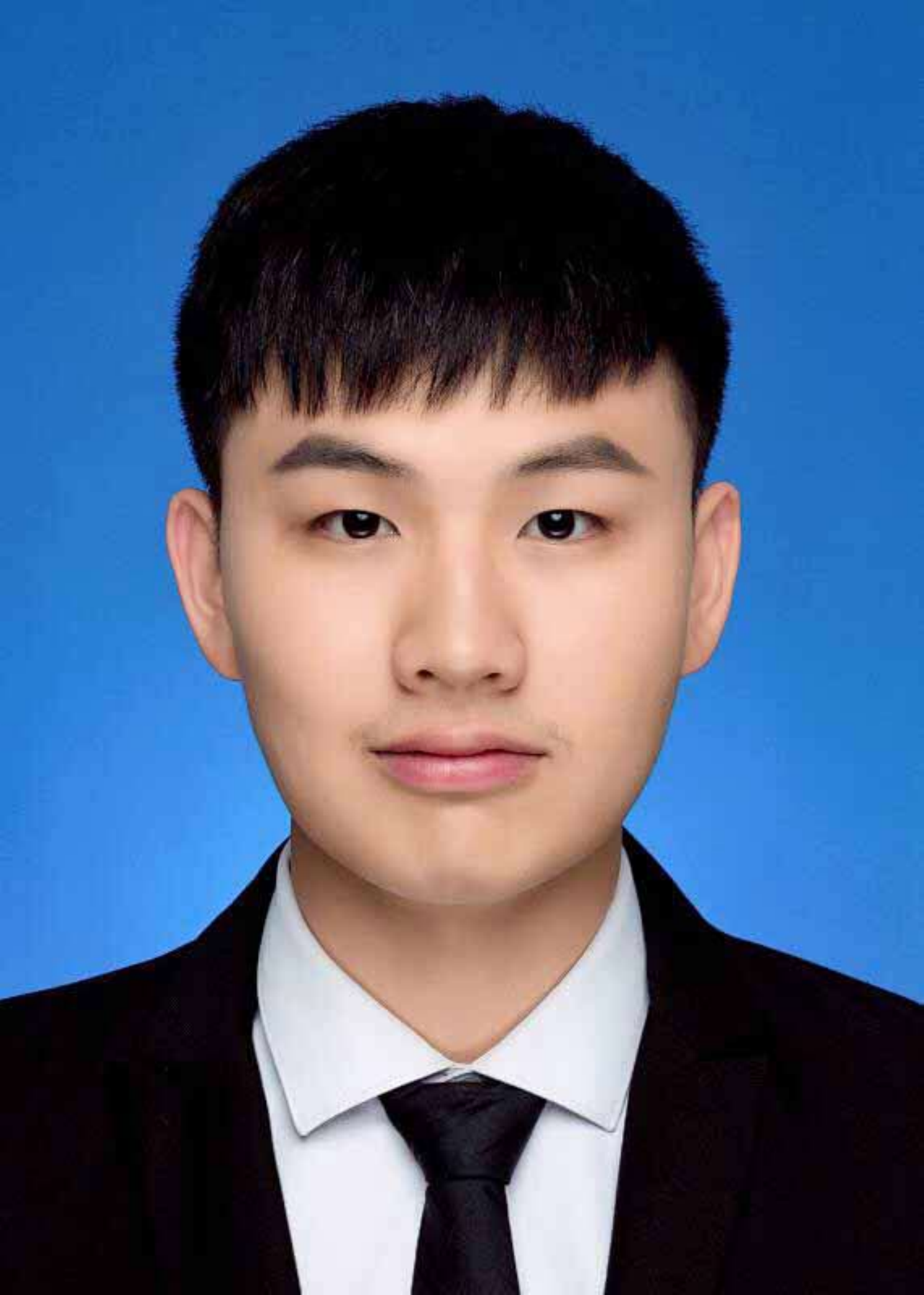}}]
{Xianzhi Zhang} received the B.S. degree from Nanchang University (NCU), Nanchang, China, in 2019. He is currently working toward the M.S. degree in Sun Yat-sen University, Guangzhou, China. His current research interests include content caching, applied machine learning and edge computing, and multimedia  communication.
\end{IEEEbiography}

\begin{IEEEbiography}[{\includegraphics[width=1in,height=1.25in,clip,keepaspectratio]{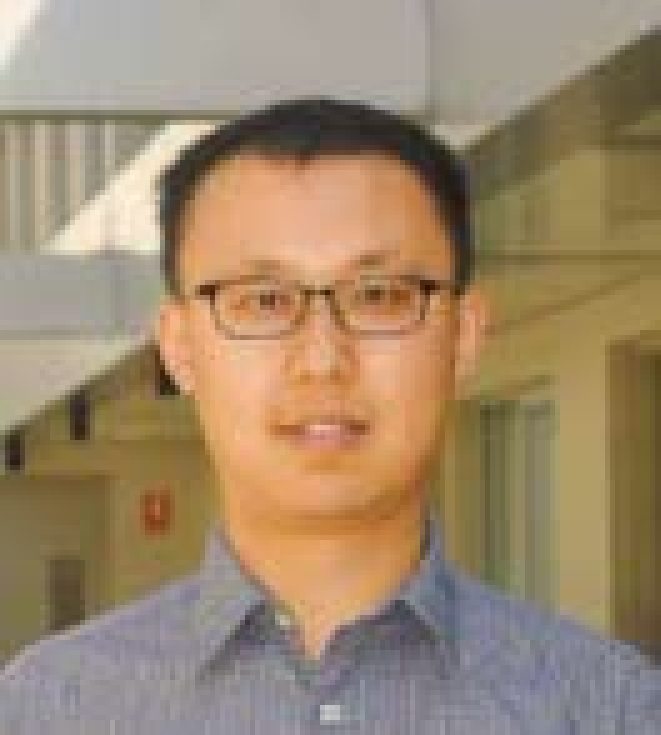}}]
{Yipeng Zhou} is a lecturer in computer science with Department of Computing at Macquarie University, and the recipient of ARC Discovery Early Career Research Award, 2018. From Aug. 2016 to Feb. 2018, he was a research fellow at the Institute for Telecommunications Research (ITR) with University of South Australia. From 2013.9-2016.9, He was a lecturer with College of Computer Science and Software Engineering, Shenzhen University. He was a Postdoctoral Fellow with Institute of Network Coding (INC) of The Chinese University of Hong Kong (CUHK) from Aug. 2012 to Aug. 2013. He won his PhD degree supervised by Prof. Dah Ming Chiu and Mphil degree supervised by Prof. Dah Ming Chiu and Prof. John C.S. Lui from Information Engineering (IE) Department of CUHK. He got Bachelor degree in Computer Science from University of Science and Technology of China (USTC).
\end{IEEEbiography}

\begin{IEEEbiography}[{\includegraphics[width=1in,height=1.25in,clip,keepaspectratio]{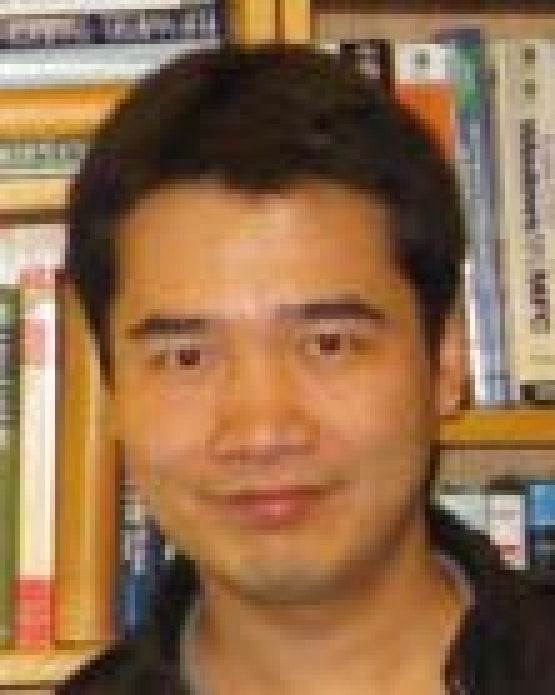}}]
{Di Wu} (M'06-SM'17) received the B.S. degree from the University of Science and Technology of China, Hefei, China, in 2000, the M.S. degree from the Institute of Computing Technology, Chinese Academy of Sciences, Beijing, China, in 2003, and the Ph.D. degree in computer science and engineering from the Chinese University of Hong Kong, Hong Kong, in 2007. He was a Post-Doctoral Researcher with the Department of Computer Science and Engineering, Polytechnic Institute of New York University, Brooklyn, NY, USA, from 2007 to 2009, advised by Prof. K. W. Ross. Dr. Wu is currently a Professor and the Associate Dean of the School of Computer Science and Engineering with Sun Yat-sen University, Guangzhou, China. His research interests include edge/cloud computing, multimedia communication, Internet measurement, and network security. 
He was the recipient of the IEEE INFOCOM 2009 Best Paper Award, IEEE Jack Neubauer Memorial Award, and etc. 
He has served as an Editor of the Journal of Telecommunication Systems (Springer), the Journal of Communications and Networks, Peer-to-Peer Networking and Applications (Springer), Security and Communication Networks (Wiley), and the KSII Transactions on Internet and Information Systems, and a Guest Editor of the IEEE Transactions on Circuits and Systems for Video Technology.
He has also served as the MSIG Chair of the Multimedia Communications Technical Committee in the IEEE Communications Society from 2014 to 2016. He served as the TPC Co-Chair of the IEEE Global Communications Conference - Cloud Computing Systems, and Networks, and Applications in 2014, the Chair of the CCF Young Computer Scientists and Engineers Forum - Guangzhou from 2014 to 2015, and a member of the Council of China Computer Federation.

\end{IEEEbiography}

\begin{IEEEbiography}[{\includegraphics[width=1in,height=1.25in,clip,keepaspectratio]{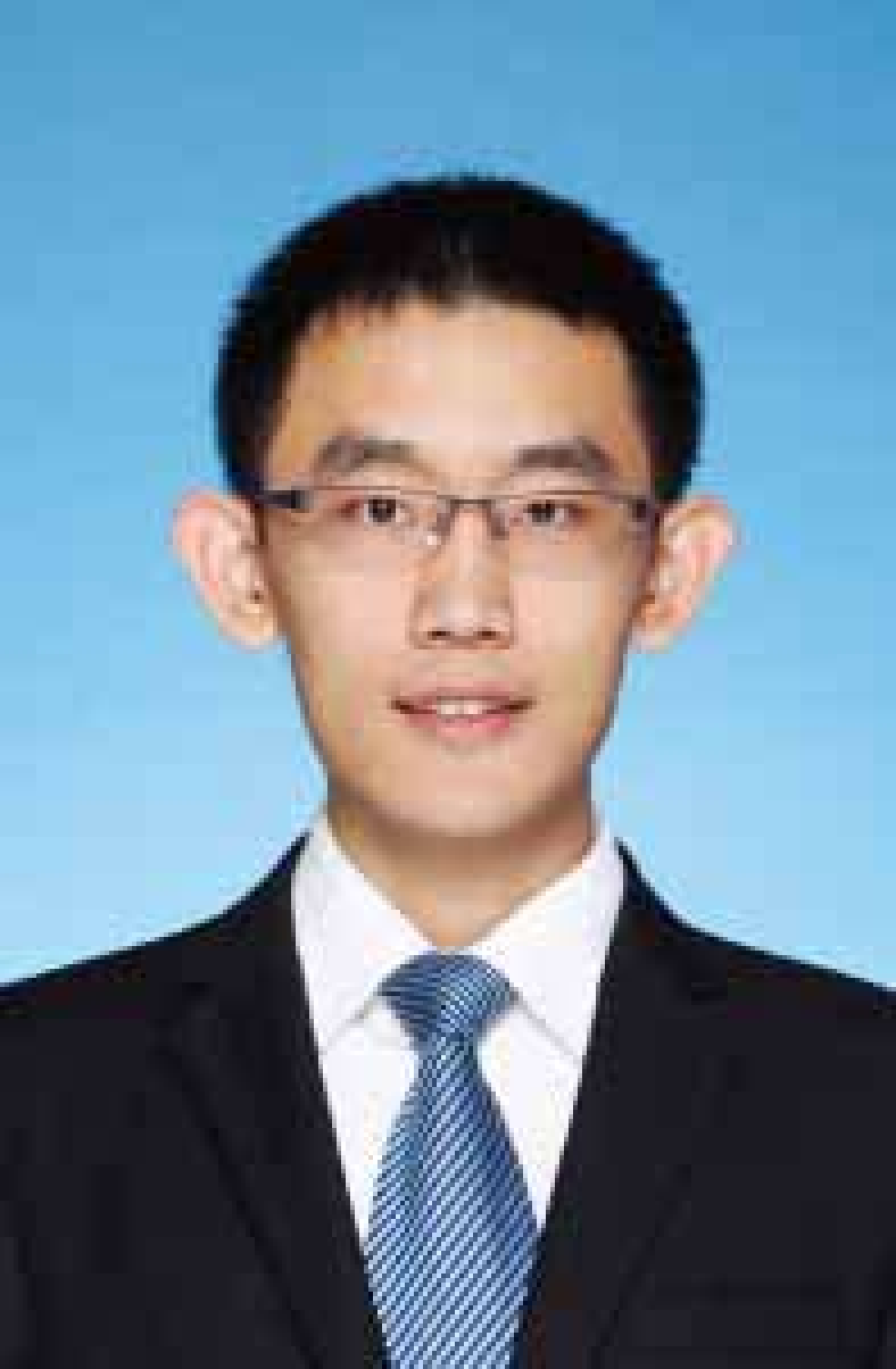}}]
{Miao Hu} (S'13-M'17) is currently an Associate Research Fellow with the School of Computer Science and Engineering, Sun Yat-Sen University, Guangzhou, China. He received the B.S. degree and the Ph.D. degree in communication engineering from Beijing Jiaotong University, Beijing, China, in 2011 and 2017, respectively. From Sept. 2014 to Sept. 2015, he was a Visiting Scholar with the Pennsylvania State University, PA, USA. His research interests include edge/cloud computing, multimedia communication and software defined networks.
\end{IEEEbiography}
\begin{IEEEbiography}[{\includegraphics[width=1in,height=1.25in,clip,keepaspectratio]{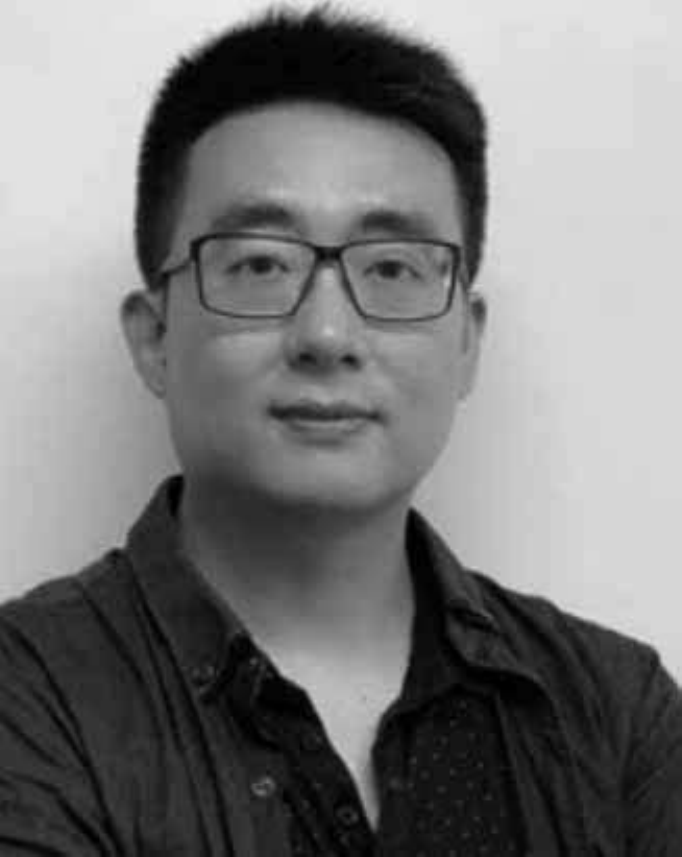}}]
{James Xi Zheng}, PhD in Software Engineering from UT Austin, Master in Computer and Information Science from UNSW,  Bachelor in Computer Information System from FuDan;  Chief Solution Architect for Menulog  Australia, now director of Intelligent systems research center (itseg.org), deputy director of software engineering, global engagement, and assistant professor in Software Engineering at Macquarie University. Specialised in Service Computing, IoT Security and Reliability Analysis.  Published more than 80  high quality publications in top journals and conferences (PerCOM, ICSE, IEEE Communications Surveys and Tutorials, IEEE Transactions on Cybernetics, IEEE Transactions on Industrial Informatics, IEEE Transactions on Vehicular Technology, IEEE IoT journal, ACM Transactions on Embedded Computing Systems). Awarded the best paper in Australian distributed computing and doctoral conference in 2017.  Awarded Deakin Research outstanding award in 2016.  His paper is recognized as a top 20 most read paper (2017-2018) in Concurrency and Computation: Practice and Experience. His another paper on IoT network security (2018) is recognized as highly cited paper.  Guest Editor and PC members for top journals and conferences (IEEE Transactions on Industry Informatics, Future Generation Computer Systems, PerCOM). WiP Chair for PerCOM 2020 and Track Chair for CloudCOM 2019. Publication Chair for ACSW 2019 and reviewers for many Trans journals and CCF A/CORE A* conferences.
\end{IEEEbiography}
\begin{IEEEbiography}
[{\includegraphics[width=1in,height=1.25in,clip,keepaspectratio]{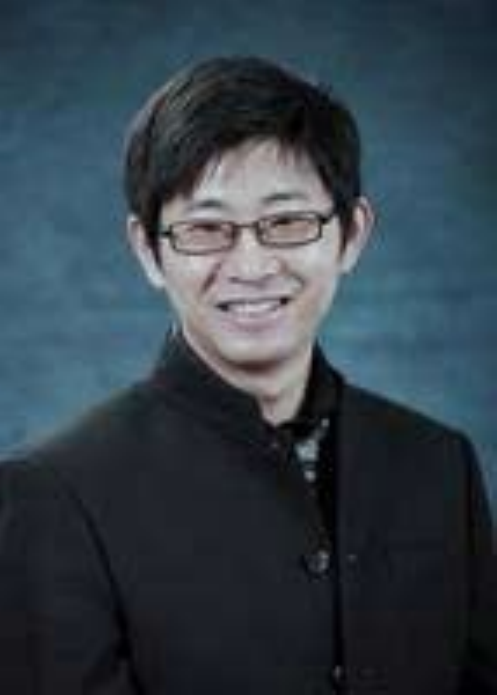}}]
{Min Chen} is a full professor in School of Computer Science and Technology at Huazhong University of Science and Technology (HUST) since Feb. 2012. He is the director of Embedded and Pervasive Computing (EPIC) Lab at HUST. He is Chair of IEEE Computer Society (CS) Special Technical Communities (STC) on Big Data. He was an assistant professor in School of Computer Science and Engineering at Seoul National University (SNU). He worked as a Post-Doctoral Fellow in Department of Electrical and Computer Engineering at University of British Columbia (UBC) for three years. Before joining UBC, he was a Post-Doctoral Fellow at SNU for one and half years. He received Best Paper Awardfrom QShine 2008, IEEE ICC 2012, ICST IndustrialIoT 2016, and IEEE IWCMC 2016. He serves as editor or associate editor for Information Sciences, Information Fusion, and IEEE Access, etc. He is a Guest Editor for IEEE Network, IEEE Wireless Communications, and IEEE Trans. Service Computing, etc. He is Co-Chair of IEEE ICC 2012-Communications Theory Symposium, and Co-Chair of IEEE ICC 2013-Wireless Networks Symposium. He is General Co-Chair for IEEE CIT-2012, Tridentcom 2014, Mobimedia 2015, and Tridentcom 2017. He is Keynote Speaker for CyberC 2012, Mobiquitous 2012, Cloudcomp 2015, IndustrialIoT 2016, and The 7th Brainstorming Workshop on 5G Wireless. He has more than 300 paper publications, including 200+ SCI papers, 80+ IEEE Trans./Journal papers, 18 ISI highly cited papers and 8 hot papers. He has published four books: OPNET IoT Simulation (2015), Big Data Inspiration (2015), 5G Software Defined Networks (2016) and Introduction to Cognitive Computing (2017) with HUST Presss, a book on big data: Big Data Related Technologies (2014) and a book on 5G: Cloud Based 5G Wireless Networks (2016) with Springer Series in Computer Science. His latest book (co-authored with Prof. Kai Hwang), entitled Big Data Analytics for Cloud/IoT and Cognitive Computing (Wiley, U.K.) appears in May 2017. His Google Scholars Citations reached 11,300+ with an h-index of 53. His top paper was cited 1100+ times. He is an IEEE Senior Member since 2009. He got IEEE Communications Society Fred W. Ellersick Prize in 2017.
\end{IEEEbiography}
\begin{IEEEbiography}
[{\includegraphics[width=1in,height=1.25in,clip,keepaspectratio]{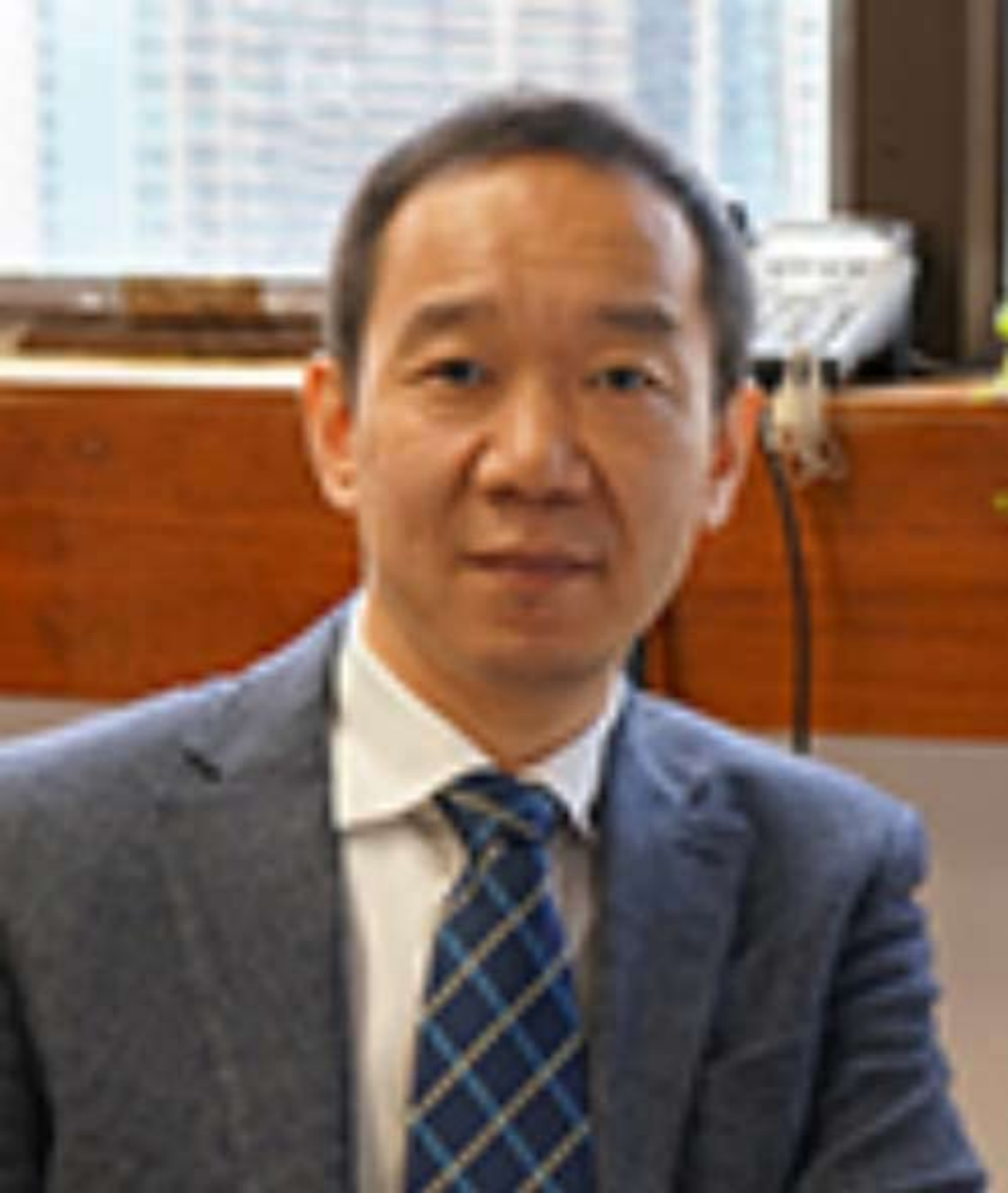}}]
{Song Guo} is a Full Professor and Associate Head (Research \& Development) in the Department of Computing at The Hong Kong Polytechnic University. He also holds a Changjiang Chair Professorship awarded by the Ministry of Education of China. Prof. Guo is an IEEE Fellow (Computer Society), a Highly Cited Researcher (Clarivate Web of Science), and an ACM Distinguished Member.
His research interests are mainly in the areas of big data, edge AI, mobile computing, and distributed systems. He co-authored 4 books, co-edited 7 books, and published over 500 papers in major journals and conferences. He is the recipient of the 2019 IEEE TCBD Best Conference Paper Award, 2018 IEEE TCGCC Best Magazine Paper Award, 2019 \& 2017 IEEE Systems Journal Annual Best Paper Award, and other 8 Best Paper Awards from IEEE/ACM conferences. His work was also recognized by the 2016 Annual Best of Computing: Notable Books and Articles in Computing in ACM Computing Reviews. Prof. Guo's research has been sponsored by RGC, NSFC, MOST, JSPS, industry, etc.
Prof. Guo is the Editor-in-Chief of IEEE Open Journal of the Computer Society and the Chair of IEEE Communications Society (ComSoc) Space and Satellite Communications Technical Committee. He was an IEEE ComSoc Distinguished Lecturer and a member of IEEE ComSoc Board of Governors. He has also served for IEEE Computer Society on Fellow Evaluation Committee, Transactions Operations Committee, Editor-in-Chief Search Committee, etc.  Prof. Guo has been named on editorial board of a number of prestigious international journals like IEEE Transactions on Parallel and Distributed Systems, IEEE Transactions on Cloud Computing, IEEE Transactions on Emerging Topics in Computing, etc. He has also served as chairs of organizing and technical committees of many international conferences.
\end{IEEEbiography}
\end{document}